\documentclass[traditabstract]{aa}
\usepackage{txfonts}
\usepackage{graphicx}
\usepackage{float}
\usepackage{natbib}
\usepackage{hyperref}
\usepackage[english]{babel}
\usepackage{color}

\title{Clustering, host halos and environment of z$\sim$2 galaxies as a function of their physical properties}

\author{Matthieu B{\'e}thermin\inst{1,2} \and Martin Kilbinger\inst{1} \and Emanuele Daddi\inst{1} \and Jared Gabor\inst{1} \and Alexis Finoguenov\inst{3} \and Henry McCracken\inst{4} \and Melody Wolk\inst{4} \and Herv\'e Aussel\inst{1} \and Veronica Strazzulo\inst{1} \and  Emeric Le Floc'h\inst{1} \and Rapha\"el Gobat\inst{1} \and Giulia Rodighiero\inst{5} \and Mark Dickinson\inst{6} \and Lingyu Wang\inst{7} \and Dieter Lutz\inst{8} \and S\'ebastien Heinis\inst{9}}

\institute{Laboratoire AIM-Paris-Saclay, CEA/DSM/Irfu - CNRS - Universit\'e Paris Diderot, CEA-Saclay, pt courrier 131, F-91191 Gif-sur-Yvette, France \and 
European Southern Observatory, Karl-Schwarzschild-Str. 2, 85748 Garching, Germany, \email{matthieu.bethermin@eso.org} \and 
Department of Physics, University of Helsinki, Gustaf H\"allstr\"omin katu 2a, FI-00014 Helsinki, Finland \and
Institut d'Astrophysique de Paris, UMR7095 CNRS, Universit\'e Pierre et Marie Curie, 98 bis Boulevard Arago, 75014 Paris, France \and
Dipartimento di Astronomia, Universita di Padova, Vicolo dell'Osservatorio 3, I-35122, Italy \and 
National Optical Astronomy Observatory, 950 North Cherry
Avenue, Tucson, AZ 85719, USA \and
Computational Cosmology, Department of Physics, University of Durham, South Road, Durham, DH1 3LE, UK \and
MPE, Postfach 1312, 85741 Garching, Germany \and
Department of Astronomy, University of Maryland, College Park, MD 20742-2421, USA}

\date{Received 17 January 2014 / Accepted 28 April 2014}

\abstract{Using a sample of 25683 star-forming and 2821 passive galaxies at $z\sim2$, selected in the COSMOS field following the BzK color criterion, we study the hosting halo mass and environment of galaxies as a function of their physical properties. \textit{Spitzer} and \textit{Herschel} allow us to obtain accurate star-formation rate estimates for starburst galaxies. We measure the auto-correlation and cross-correlation functions of various galaxy sub-samples  and infer the properties of their hosting halos using both a halo occupation model and the linear bias at large scale. We find that passive and star-forming galaxies obey a similarly rising relation between the halo and stellar mass. The mean host halo mass of star forming galaxies increases with the star formation rate between 30 and 200\,M$_\odot$.yr$^{-1}$, but flattens for higher values, except if we select only main-sequence galaxies. This reflects the expected transition from a regime of secular co-evolution of the halos and the galaxies to a regime of episodic starburst. We find similar large scale biases for main-sequence, passive, and starburst galaxies at equal stellar mass, suggesting that these populations live in halos of the same mass. However, we detect an excess of clustering on small scales for passive galaxies and showed, by measuring the large-scale bias of close pairs of passive galaxies, that this excess is caused by a small fraction ($\sim16\%$) of passive galaxies being hosted by massive halos ($\sim 3 \times 10^{13}$\,M$_\odot$) as satellites. Finally, extrapolating the growth of halos hosting the z$\sim$2 population, we show that M$_\star \sim 10^{10}$\,M$_\odot$ galaxies at z$\sim$2  will evolve, on average, into massive (M$_\star \sim 10^{11}$\,M$_\odot$), field galaxies in the local Universe and M$_\star \sim 10^{11}$\,M$_\odot$ galaxies at z=2 into local, massive, group galaxies. We also identify two z$\sim$2 populations which should end up in today's clusters: massive ($>$M$_\star \sim 10^{11}$\,M$_\odot$), strongly star-forming ($>200$\,M$_\odot$.yr$^{-1}$), main-sequence galaxies, and close pairs of massive, passive galaxies.}


\titlerunning{Clustering and host-halo mass of z$\sim$2 galaxies as a function of their physical properties}

\authorrunning{B\'ethermin et al.}

\begin{document}

\maketitle

\section{Introduction}

Understanding galaxy formation and evolution in the context of the standard $\Lambda$CDM cosmological model is one of the main challenges of modern astrophysics. For almost two decades, semi-analytical models have tried to reproduce the statistical properties of galaxies from the evolution of dark matter structures and using analytical recipes for baryonic physics calibrated on hydrodynamical simulations \citep[e.g.][]{Guiderdoni1998,Somerville1999,Hatton2003}. However, these models are not able to accurately reproduce  the infrared and submillimeter number counts of galaxies, which directly probe  star formation in  galaxies at high redshift, without invoking strong assumptions like
advocating the adoption of a top-heavy initial mass function (IMF) in high redshift major and minor mergers \citep{Baugh2005},  ad-hoc inefficiencies of star-formation in low-mass halos \citep{Bouche2010,Cousin2013}, or excessively long delays for the re-accretion of the gas ejected by supernovae \citep{Henriques2013}. This population of high-redshift (z$>$2), intensely star-forming (SFR$>$200\,M$_\odot$/yr) galaxies is important, because such galaxies are thought  to be the progenitors of  massive and passive galaxies present in the local Universe \citep[e.g.][]{Daddi2007,Tacconi2008,Cimatti2008}. The exact nature of the mechanism(s), which trigger the transformation of star-forming galaxies into passive elliptical galaxies  is also an open question. Feedback from active galactic nuclei (AGN) is often advocated for this \citep[e.g.][]{Cattaneo2006,Somerville2008}, but other mechanisms such as the suppression of gas cooling in the most massive halos due to their hot atmosphere  \citep[e.g.][]{Keres2005,Birnboim2007} are also possible.\\

Recent observational studies revealed interesting insights about the nature of star formation in  high-redshift massive galaxies. Measurements of the stellar mass function of galaxies showed that a significant fraction ($\sim$30\%) of massive galaxies (M$_\star>$10$^{11}$\,M$_\odot$) are already passive at z$\sim$2, when the vast majority of low-mass galaxies are star-forming \citep{Ilbert2010,Ilbert2013,Muzzin2013}. In addition, detailed studies of star-forming galaxies at these redshifts found a strong correlation between the star formation rates (SFRs) and stellar masses \citep[e.g.][]{Daddi2007,Rodighiero2010}, the so-called main sequence, which suggests that the star-formation is driven by universal, secular processes. \textit{Herschel} showed that a few percent of the massive star-forming galaxies are strong outliers of this sequence and present an excess of specific star formation rate (sSFR=SFR/M$_\star$) by a factor of at least four compared to the main sequence  \citep{Elbaz2011,Rodighiero2011,Sargent2012}. These episodic starbursts are probably induced by major mergers \citep{Daddi2010b,Hung2013}. However, the strong diversity of the star-formation properties in galaxies at high redshift is not well understood. It could be in principle related to the properties of the host dark matter halos, hence to environmental effects. This can be investigated by measuring the clustering of galaxies of different types. For instance, the clustering of high-redshift starbursts can discriminate between major-merger-driven and secular star-formation processes \citep{VanKampen2005}.\\

Because of the invisible nature of the dark matter, measuring the hosting halo mass of a galaxy sample is difficult. Nevertheless, the link between halo mass (M$_h$) and stellar mass (M$_\star$) was studied extensively during the last decade. The halo occupation modeling allows us to infer how galaxies are distributed inside halos from observations of their clustering \citep[for a review]{Cooray2002}. Using this technique, \citet{Coupon2012} measured the relation between halo and stellar mass up to z=1.2. A slightly different but complementary approach is the abundance matching technique, which connects halo mass and stellar mass of galaxies directly from the related mass functions, assuming a monotonic relation between these two quantities \citep[e.g.][]{Vale2004}. Finally,  weak gravitational lensing can also provide strong constraints on the characteristic halo mass hosting a galaxy population \citep[e.g.][]{Mandelbaum2006}. \citet{Leauthaud2012} made a combined analysis of z$<$1 galaxies combining all these technique and strongly constrained the M$_\star$-M$_h$ relation. However, only a few studies extended these results at higher redshift. Among these, there are studies based on abundance matching going up to z=4 by \citet{Behroozi2010} and \citet{Moster2010} and a work based on abundance matching and clustering at z$<$2 by \citet{Wang2012}. Recently, Wolk et al. (in prep.) pushed the studies of the M$_\star$-M$_h$ relation up to z$\sim$2.5 and lower stellar masses using both clustering and abundance  matching with the UltraVISTA data \citep{McCracken2012}.\\

The link between halo mass and other properties like the star formation rate (SFR) or specific star formation rate (sSFR=SFR/M$_\star$) has been less explored. However, some interesting analyses were recently performed. Empirical models \citep[e.g.][]{Conroy2009,Behroozi2013} calibrated on the evolution of stellar mass function applied a simple prescriptions to estimate the mean relation between M$_h$ and SFR. Some other empirical models used the link between M$_\star$ and SFR estimated from  UV and far-infrared observations and the well-studied M$_h$-M$_\star$ relation to determine in turn the link between SFR and M$_h$ \citep[e.g.][]{Bethermin2012a,Wang2012,Bethermin2013}. \citet{Lee2009} studied the UV-light-to-halo-mass ratio at $3<z<5$, and found a decrease of this ratio with time at fixed halo mass. Finally, \citet{Lin2012} studied the clustering of z$\sim$2 galaxies as a function of SFR and sSFR, estimated using UV luminosity corrected for dust extinction, and found  clustering   increasing with the distance of galaxies from the main sequence (i.e., with sSFR). Finally, several studies using the correlated anisotropies of the cosmic infrared background (CIB, which is the relic emission of the dust emission from all star-forming galaxies across cosmic times) showed that from z=0 to z=3 the bulk of the star formation is hosted in majority by halos of $\sim$10$^{12-13}$\,M$_\odot$ \citep[e.g.][]{Bethermin2013,Viero2013,Planck_CIB_2013}. In particular, \citet{Bethermin2013} showed that the strong evolution of populations of star-forming galaxies responsible for the CIB can be modeled assuming an universal efficiency of conversion of accreted cosmological gas into stars as a function of the halo mass peaking around 10$^{12.5}$\,M$_\odot$ and the evolution of the accretion rate at fixed halo mass with time.\\

In this paper, we study the clustering of individually-detected galaxy populations focusing on the 1.5$<$z$<$2.5 redshift range, when star formation is maximal \citep[e.g.][]{Hopkins2006,Le_Borgne2009,Gruppioni2013,Magnelli2013,Burgarella2013,Planck_CIB_2013} to obtain new observational constraints on the link between the star-forming properties of galaxies and the nature of their host halos and environments. Ultimately, our aim is a better understanding of the mechanisms which drive star formation, trigger starbursts and quench galaxies at high redshift.\\

In Sect.\,\ref{sect:sample}, we describe the approach used to build our sample and the estimate of the physical properties of galaxies. In Sect.\,\ref{sect:formalism}, we detail the method chosen to measure the angular correlation function of our various sub-samples and the halo occupation model used to interpret the measurements. Sect.\,\ref{sect:sample} and Sect.\,\ref{sect:formalism} can be skipped by readers not interested in the technical details. In Sect.\,\ref{sect:mstarmh}, \ref{sect:sfrmh}, and \ref{sect:ssfrmh}, we present and discuss our results on the link between the halo mass and stellar mass, the SFR, and the sSFR, respectively. In Sect.\,\ref{sect:cross}, we study the clustering for galaxies depending of their nature, i.e., for the categories main-sequence, starburst, or passive. In Sect.\,\ref{sect:discussion}, we discuss the consequences of our results on our understanding of galaxy evolution. We finally conclude in Sect.\,\ref{sect:conclusion}.\\

In this paper, we assume a WMAP-7 cosmology \citep{Larson2010} and a \citet{Salpeter1955} initial mass function. We use the virial mass for the definition of the dark matter halo mass.\\

\section{Description of the sample}

\label{sect:sample}

We built a sample of galaxies at z$\sim$2 to perform our analysis. We used the BzK selection technique of \citet{Daddi2004}, which allows to  efficiently select galaxies around z=2, to split  the sample into a star-forming and a passive galaxy population, and to estimate the stellar mass and the SFR  using only B, z, and K-band photometry (see \citealt{Daddi2004,Daddi2007} for details). Stellar masses and UV-based SFRs computed in this way have formal errors typically around 0.1--0.2~dex or lower, and within a 0.3~dex scatter are in agreement with those computed from the fit of global SEDs (see also e.g. \citealt{Rodighiero2014}). We use the same band-merged photometric catalog as in \citet{McCracken2010}, selected down to K$_{AB}=23$. However, the UV-derived SFR is not reliable for dust-obscured starbursts \citep{Goldader2002,Chapman2005,Daddi2007}. We thus used \textit{Spitzer} and \textit{Herschel}-derived SFR, where available. This "ladder of SFR indicators" is  similar to the one built by \citet{Wuyts2011a} and to the one used by \citet{Rodighiero2011}.\\

\subsection{COSMOS passive BzK sample}

The high-redshift passive galaxies (called hereafter pBzK) are selected using the following criteria \citep{Daddi2004}:
\begin{equation}
(z_{\rm AB}-K_{\rm AB}) - (B_{\rm AB}-z_{\rm AB}) \leq -0.2 \ \textrm{and} \ (z-K)_{\rm AB} > 2.5,
\end{equation}
where z$_{\rm AB}$, K$_{\rm AB}$ and B$_{\rm AB}$ are the magnitude in AB convention of the galaxies in the corresponding bands. To avoid any contamination by low-z interlopers, we also discard galaxies with a photometric redshift (coming from \citealt{Ilbert2009}) lower than 1.4. The stellar mass is estimated from the K-band photometry and using the $z-K$ color to estimate the mass-to-light ratio following \citet{Daddi2004}. The stellar mass is only weakly dependent on the exact redshift of the sources for pBzKs and no correction taking into the photometric redshift is performed.

\subsection{COSMOS star-forming BzK sample}

The high-redshift, star-forming galaxies (called hereafter sBzK) lie in another part of the BzK diagram \citep{Daddi2004}:
\begin{equation}
(z_{\rm AB}-K_{\rm AB}) - (B_{\rm AB}-z_{\rm AB}) \geq -0.2.
\end{equation}
We apply the same cut on the photometric redshift (z$>$1.4) to remove the low-redshift interlopers and use the same method as for pBzK to estimate their stellar mass. The SFR in sBzK is estimated from the B-band photometry, which is corrected for attenuation estimating the UV-slope using the (B-z) color. The SFR estimate is more sensitive to the exact redshift of the source in the interval, and the photometric redshift is used to refine the value of SFR when it is available (95\% of the sample). In addition to these criterion, we remove the objects which are classified as passive in the UVJ diagram (\citealt{Williams2009}, see also \citealt{Wuyts2007}) from the sBzK sample.\\

\subsection{\textit{Spitzer}/MIPS data}

In highly-obscured, dusty galaxies, only a small fraction of UV light can escape from the interstellar dust clouds hosting young stars and UV-corrected estimates of SFR are not reliable (see discussion in \citealt{Rodighiero2011}). An estimate of SFR from infrared data is then more reliable. \citet{LeFloch2009} built a catalog of 24\,$\mu$m sources matched with the data from which the BzK sample was built. The total infrared luminosity (L$_{\rm IR}$, integrated between 8 and 1000\,$\mu$m) is extrapolated from the 24\,$\mu$m flux density using the \citet{Magdis2012} templates. L$_{\rm IR}$ is then converted into SFR assuming the \citet{Kennicutt1998} conversion factor.\\

\subsection{\textit{Herschel}/PACS data}

At very high 24\,$\mu$m flux density, the 24-$\mu$m-derived SFR is also no longer reliable. On the one hand, AGN contamination becomes significant \citep[e.g.][]{Treister2006}. On the other hand, the ratio between PAH features and the cold dust continuum is lower in starbursting galaxies \citep{Elbaz2011}. For this reason, we use the \textit{Herschel}/PACS catalog of the COSMOS field, which was extracted using the positions of 24\,$\mu$m sources as a prior. The data comes from the PACS evolutionary probe survey (PEP, \citealt{Lutz2011}). We fitted the 100 and 160\,$\mu$m PACS flux densities with \citet{Magdis2012} templates in order to derive L$_\textrm{IR}$ of each galaxy and assume the same conversion factor between L$_{\rm IR}$ and SFR as for MIPS. PACS wavelengths have the advantage of being close to the maximum of emission of the dust and the recovered L$_{\rm IR}$ is few sensitive to the assumed temperature.\\

\subsection{SFR-M$_\star$ diagram}

\begin{figure}
\centering
\includegraphics{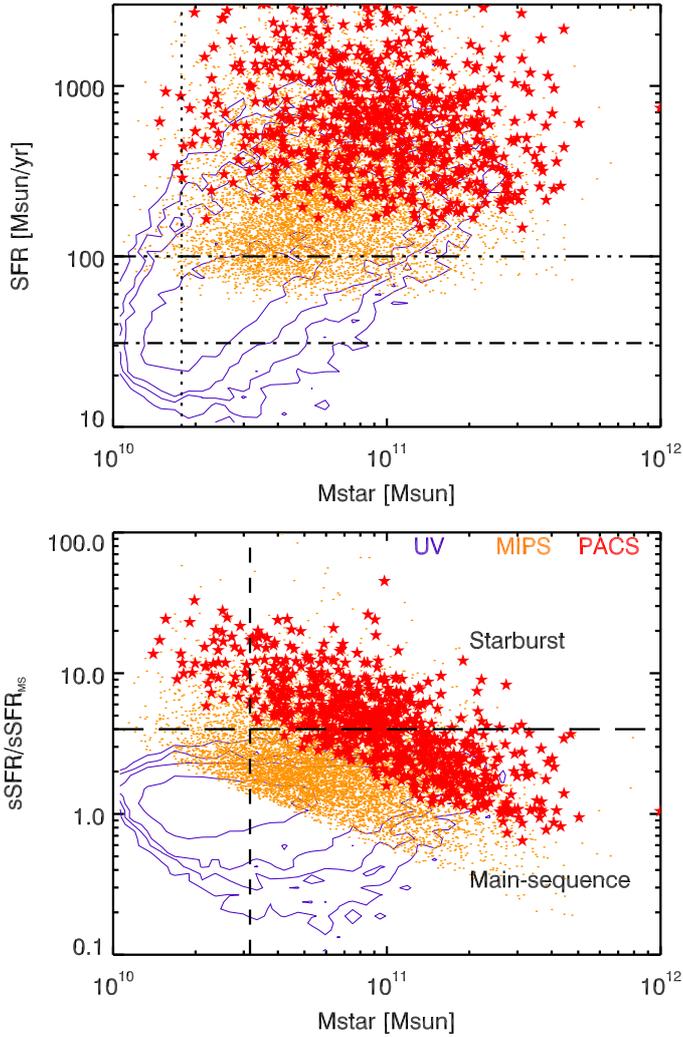}
\caption{\label{fig:ms} \textit{Upper panel:} Position of our sBzK sources in the SFR-M$_\star$ plane. The purple contours indicate sources with only UV-detection (i.e., having no mid/far-IR), the orange dots the MIPS-but-not-PACS-detected sources, and the red stars the PACS-detected sources. The vertical dotted line indicates the stellar mass where the sample becomes strongly incomplete. The horizontal dot-dash and three-dot-dash lines represents the completeness limits in SFR of the UV and MIPS sub-samples. \textit{Lower panel:} Distance between our sBzK sources and the center of the main-sequence. The horizontal long-dash line corresponds to the limit between the galaxies classified as main sequence of starburst. The vertical short-dash line shows the mass cut that defines a complete mass-selected sample of starburst galaxies detected by PACS.}
\end{figure}

Fig.\,\ref{fig:ms} (upper panel) shows the position of our sBzK sources in the classical SFR-M$_\star$ diagram. The correlation (so-called main sequence) between SFR and M$_\star$ is well probed using UV-derived SFRs (purple contours). The pBzKs, not showed in the figure, have a low star formation, which is thus very difficult to estimate accurately. They lie well below the main-sequence. Because of their high threshold in SFR, the correlation is poorly detected by MIPS (yellow dots) and not seen by PACS (red stars). This shows why it is so important to use various SFR estimators to accurately probe the full SFR-M$_\star$ diagram. In this paper, we will present the clustering as a function of various physical parameters (M$_\star$, SFR, sSFR). We have to define various cuts for which our samples are complete.\\

Below M$_\star$=10$^{10.25}$\,M$_\odot$, the catalog of star-forming BzK becomes incomplete. This is caused by the magnitude limits of the catalog in the B, z, and K bands. Consequently, we will apply these cuts, when we study the clustering as a function of stellar mass. For pBzK galaxies, the catalog begins to be incomplete below M$_\star \sim 10^{10.8}$\,M$_\odot$. We define two sample of which one highly complete above $10^{11} \, M_\odot$ and we also use a mass bin between 10$^{10.5}$ and 10$^{11}$ \,M$_\odot$, albeit somewhat incomplete. Incompleteness is not a substantial problem for clustering studies unless it is correlated with environment. Concerning the selections in SFR, we wish to define an SFR limit above which the sample is not affected by the mass incompleteness caused by the K-band sensitivity limit. We estimated this SFR cut to be 30\,M$_\odot$.yr$^{-1}$ (see Fig.\,\ref{fig:ms} lower panel). MIPS and PACS are only sensitive to SFR $\gtrsim$100\,M$_\odot$/yr and $\gtrsim$200\,M$_\odot$.yr$^{-1}$, respectively, and are not affected by the incompleteness in mass. For MIPS, we use a sharp cut at 100\,M$_\odot$.yr$^{-1}$. For the PACS selected sample, we use the full sample to maximize the statistics, because the total number of detections is already small for a clustering study.\\

The selection in sSFR is more tricky. The same sSFR can be measured in a low mass galaxy hosting a low SFR and in a massive, strongly-star-forming galaxies. It is then impossible to define a completeness cut for sSFR. For this reason, we first apply a stellar mass threshold before sorting the galaxies by sSFR. We have chosen a slightly high cut of 10$^{10.5}$\,M$_\odot$, which allows to detect all the M$_\star > $10$^{10.5}$\,M$_\odot$ starburst galaxies with PACS. The starburst galaxies are defined as being 0.6\,dex (a factor of 4) above the main-sequence following \citet{Rodighiero2011}. There is no clear gap between main-sequence and starburst galaxies in the sSFR distribution. However, \citet{Rodighiero2011} and \citet{Sargent2012} showed evidence for a departure from a log-normal distribution at sSFRs 0.6\,dex larger than the center of the main-sequence. Modeling studies \citep{Bethermin2012b,Sargent2013} suggested that the value of 0.6\,dex corresponds to the transition between secularly-star-forming galaxies and merger-induced starbursts. Nevertheless, this bimodality of the star-formation modes does not imply a strict separation in the SFR-M$_\star$ diagram. The PACS detection is crucial in our analysis, because starbursts are highly obscured, which makes the UV-derived SFR not very reliable. In addition, they present a PAH deficit \citep{Elbaz2011}, which implies an underestimated 24\,$\mu$m-derived SFR. Fig.\,\ref{fig:ms} (lower panel) shows the distance between the galaxies and the main sequence. We use the same definition as \citet{Bethermin2012c} for the center of the main-sequence:
\begin{equation}
\textrm{sSFR}_{MS} = 10^{-10.2} \textrm{yr}^{-1} \times \left ( \frac{M_\star}{10^{11}\,M_\odot} \right)^{-0.2} ( 1+ z)^3,
\end{equation}
where z and M$_\star$ are the best estimate of the photometric redshift \citep{Ilbert2010} and the stellar mass for each galaxy.

\subsection{Redshift and mass distributions}

\label{sect:Nz}

\begin{figure}
\centering
\includegraphics{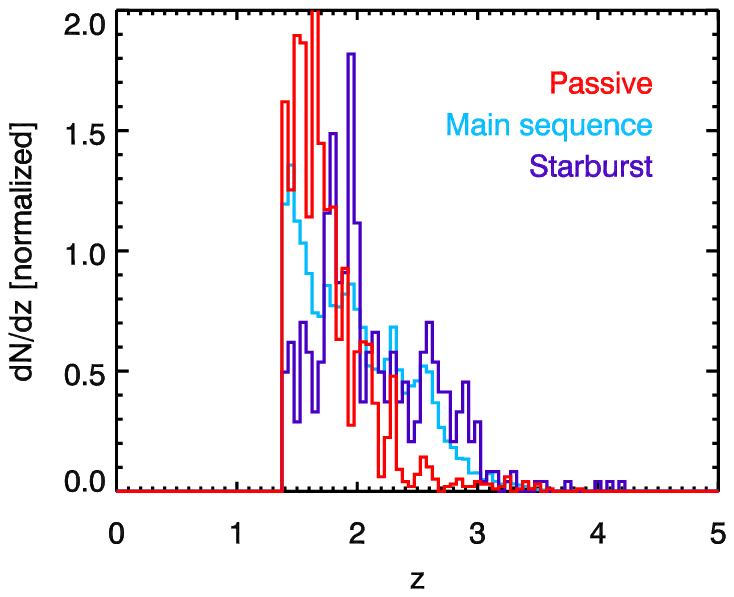}
\includegraphics{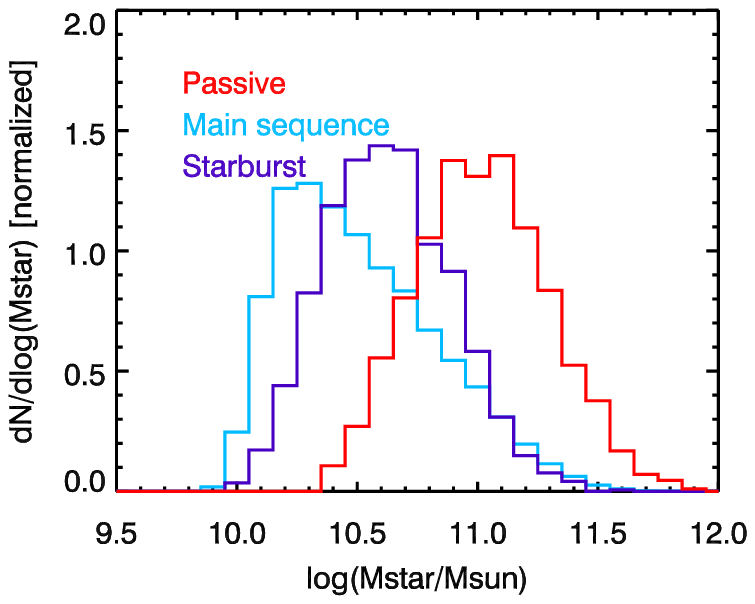}
\caption{\label{fig:Nz} \textit{Upper panel:} Redshift distribution for our samples of passive (red), main-sequence (light blue), and starburst (purple) galaxies. \textit{Lower panel:} Stellar-mass distributions.}
\end{figure}

In order to interpret the clustering measurements, we need to know the redshift distribution of our sources. Figure\,\ref{fig:Nz} (upper panel) shows the redshift distribution of our various sub-samples, i.e. pBzK-selected passive galaxies, sBzK-selected main-sequence galaxies, and sBzK-and-PACS-selected starbursts. For the three categories, the bulk of the source lies between z=1.4 and z=2.5. However, the star-forming galaxies have a tail at z$>$2.5, but not the passive galaxies, probably because a small fraction of galaxies, including the massive ones, are already quenched at z$>$2 \citep{Ilbert2013,Muzzin2013}, and also likely due to photometric selection effects.\\ 

The distribution of stellar mass of the three sub-samples is shown Fig.\,\ref{fig:Nz} (lower panel). The rapid decrease of the number of main sequence and starburst galaxies below $\sim 10^{10}$\,M$_\odot$ is caused by the magnitude limit of the survey in band K. The small offset ($\sim$0.2\,dex) between the position of this break for this two population is probably caused by a larger attenuation in the starbursts, and the FIR depth required to select starbursts. The mass distribution of passive galaxies peaks around 10$^{11}$\,M$_\odot$ in agreement with \citet{Ilbert2013} and \citet{Muzzin2013}, but we caution that in our sample this happens fairly close to the completeness limit.\\

\section{Clustering measurements and modeling}

\label{sect:formalism}

\subsection{Measurements of the correlation function}

One of the simplest estimators of the clustering of galaxies is the auto-correlation function (ACF), written often as $w(\theta)$. This function is the excess probability over Poisson  of finding a neighbor at an angular distance $\theta$ from another source. This function can be measured using the \citet{Landy1993} estimator:
\begin{equation}
w(\theta) = \frac{1}{RR} \times \left ( \frac{N_{\rm r} (N_{\rm r}-1)}{N_{\rm d} (N_{\rm d}-1)} DD - 2 \frac{N_{\rm r}}{N_{\rm d}} DR + RR \right ),
\end{equation}
where $N_{\rm r}$ and $N_{\rm d}$ are the number of objects in the random and galaxy catalogs, respectively. DD, DR and RR are the number of pairs, separated by an angle between $\theta - d\theta/2$ and $\theta + d\theta/2$, in the galaxy catalog, between the random and the galaxy catalogs, and in the random catalog, respectively. The random catalog contains 100\,000 objects to minimize the Poisson errors coming from the statistical fluctuations of DR and RR. It was drawn in an uniform way in the field excluding the masked regions. The angular, cross-correlation function $\chi_{\rm AB}$ between two population (A and B) is computed with:
\begin{equation}
\begin{split}
\chi_{\rm AB}(\theta) = \frac{1}{R_{\rm A}R_{\rm B}} \times \left ( \frac{N_{\rm rA} N_{\rm rB}}{N_{\rm A} N_{\rm B}} D_{\rm A}D_{\rm B} - \frac{N_{\rm rB}}{N_{\rm A}} D_{\rm A}R_{\rm B}   - \frac{N_{\rm rA}}{N_{\rm B}} D_{\rm B}R_{\rm A} \right ),\\
+R_{\rm A}R_{\rm B} \, \, \, \, \, \, \, \, \, \, \, \, \, \, \, \, \, 
\end{split}
\end{equation}\\
where $N_{\rm A}$ is the number of galaxies in the population A ($N_{\rm B}$ in population B), $N_{\rm rA}$ the number of sources in the random catalog A ($N_{\rm rB}$ in the random catalog B), $D_{\rm A}D_{\rm B}$ the number of pairs between the population A and B, and $D_{\rm A}R_{\rm B}$ ($D_{\rm B}R_{\rm A}$, respectively) between population A (B, respectively) and the random catalog B (A, respectively). These computations were performed with the ATHENA code\footnote{http://www2.iap.fr/users/kilbinge/athena/}. As \citet{Coupon2012}, we estimated the uncertainties using a jackknife method, splitting the field into 20 sub-fields. This technique also allows to estimate the covariance matrix between the angular bins, which is used to fit the models (see Sect.\,\ref{sect:hod}).\\

The \citet{Landy1993} estimator is biased by an offset called integral constraint $C$ ($w_{\rm mes} = w - C$, where $w_{\rm mes}$ is the measured ACF and $w$ the intrinsic one):
\begin{equation}
C = \frac{1}{\Omega_{\rm field}^2}\int_{\rm field} \int_{field} w(\theta) \, d\Omega_1 \, d\Omega_2,
\end{equation}
where $\Omega_{\rm field}$ is the size of the field and $\theta$ is the distance between two points drawn randomly in the field. The double integral takes into account all the possible random pairs in the field. In practice, we pre-computed the angles between $10^6$ pairs randomly taken in the field. Before comparing a model to the data, we compute C taking the mean $w(\theta)$ value provided by the model for these angles. The method is exactly the same for the cross-correlation function.\\

\subsection{Estimation of the correlation length}

A simple way to interpret the clustering of a population is to determine its correlation length $r_0$. We will estimate this value for our various samples to allow an easy comparison with the literature. The 3-dimension angular correlation function is assumed to be described by a power law:
\begin{equation}
\epsilon(r) = \left ( \frac{r}{r_0} \right )^{-\gamma},
\end{equation}
where $r$ is the comoving distance between the two points and $r_0$ the characteristic correlation length. $\gamma$ is fixed to the standard value of 1.8.\\

In the flat-sky approximation of \citet{Limber1953}, the corresponding angular correlation function is then \citep{Peebles1980}:
\begin{equation}
w(\theta) = \theta^{1-\gamma} r_0^\gamma \frac{\Gamma(\gamma/2)}{\Gamma(1/2) \Gamma((\gamma-1)/2)} \frac{c}{H_0} \frac{\left ( \int \frac{dN}{dz} dz \right)^2}{\int E(z) D_{\rm C}^{1-\gamma} \left ( \frac{dN}{dz} \right )^2 dz},
\end{equation}
where $H_0$ is the Hubble constant, $E(z) = \sqrt{(1+z)^3 \Omega_{\rm m} + \Omega_\Lambda}$, $D_{\rm C}$ the comoving distance corresponding to a redshift $z$, and $dN/dz$ the redshift distribution of the sample. We used the redshift distribution of passive, main sequence, or starburst galaxies presented in Sect.\,\ref{sect:Nz}. We checked that using the redshift distribution of the full parent population, i.e. all passive, all main sequence, or starburst galaxies depending on the case, or only the one of mass- or SFR-selected sub-samples, does not bias significantly ($>$1$\sigma$) the results. We thus used the redshift distributions of the full parent samples in our analysis to avoid any problems caused by low statistics for the small sub-samples. On small scales, the correlation function is not exactly a power-law and exhibits an excess caused by the correlation of galaxies inside the same dark matter halo. This is especially relevant for passive galaxies as discussed in Sect.\,\ref{sect:cross}. We thus fit only scales larger than 30\,arcsec to determine $r_0$.\\

\subsection{Determination of the mean mass of the host halos}

We also used a standard halo occupation distribution (HOD) model (see \citealt{Cooray2002} for review), described in Sect.\,\ref{sect:hod} and a method based on the large-scale clustering, explained in Sect.\,\ref{sect:lsbias}, to estimate the typical mass of halos hosting the various sub-samples we studied in this paper. The consistency of these methods and the validity of the assumptions on which they are based is discussed in Sect.\,\ref{sect:validity}.\\

\subsubsection{With a halo occupation model}

\label{sect:hod}
 
The HOD models interpret the clustering of galaxies using a parametric description of how galaxies occupy halos. We follow the same approach as \citet{Coupon2012}, based on \citet{Zheng2005} formalism. This formalism was initially built to study samples selected using a mass threshold. In Sect.\,\ref{sect:validity}, we discuss the pertinence of this formalism for our samples.\\

In the HOD formalism, the total number of galaxies in a halo of mass M$_h$ is described by
\begin{equation}
N(M_h) = N_c \times \left ( 1+ N_s(M_h) \right ),
\end{equation}
where N$_c$ is the number of central galaxies and N$_s$ the number of satellites. These two quantities are assumed to vary only with halo mass. Classically, $N_c$ is parametrized by
\begin{equation} 
N_c(M_h) = \frac{1}{2} \left ( 1+ \textrm{erf} \left ( \frac{\textrm{log}(M_h)-\textrm{log}(M_{\rm min})}{\sigma_{\rm log(M)}} \right ) \right ),
\end{equation}
where M$_{\rm min}$ is the minimal mass of halos hosting central galaxies and $\sigma_{\textrm{log}(M)}$ is the dispersion around this threshold. Another parametrization is used for N$_s$:
\begin{equation} 
N_s(M_h) = \left ( \frac{M_h-M_0}{M_1} \right )^\alpha,
\end{equation}
where M$_1$, M$_0$, and $\alpha$ are free parameters of the HOD models.\\

The auto-correlation function can be derived using the classical formula:
\begin{equation}
w(\theta) = \frac{\int_z \left ( \frac{dN}{dz} \right )^2 \int_{k} \frac{k}{2 \pi} P_{\rm gg}(k,z) J_0(k D_c \theta) \, dz \, dk}{\left ( \int_z \frac{dN}{dz}  \, dz \right )^2},
\label{eq:whod}
\end{equation}
where $J_0$ is the zero-th order Bessel function and $dN/dz$ the redshift distribution of the galaxy sample (see Sect.\,\ref{sect:Nz}). $P_{\rm gg}$ is the sum of two terms corresponding to the clustering of galaxies in two different halos and inside the same halo:
\begin{equation}
P_{\rm gg}(k,z) = P_{\rm gg}^{\rm 2h}(k,z) + P_{\rm gg}^{\rm 1h}(k,z).
\end{equation}
Following the standard method, we compute the 2-halo term with
\begin{equation}
P_{\rm gg}^{\rm 2h}(k,z) =  \left [ \int_{M_h} \frac{d^2N}{d\textrm{log}(M_h) dV} b(M_h) \frac{N_c + N_s}{\bar{n}_{\rm gal}} dM_h\right ]^2 P_{\rm lin}(k,z),
\label{eq:hod2h}
\end{equation}
where $d^2N/d\textrm{log}(M_h) dV$ is the halo mass function and $\bar{n}_{\rm gal} = \int  \frac{d^2N}{d\textrm{log}(M_h) dV} (N_c +N_s) dM_h$ the mean number of galaxies. The 1-halo term is computed with
\begin{equation}
P_{\rm gg}^{\rm 1h}(k,z) =  \int_{M_h} \frac{d^2N}{d\textrm{log}(M_h) dV} \frac{2 N_c N_s + N_s^2}{\bar{n}_{\rm gal}^2} u^2(k,M_h,z) dM_h,
\end{equation}
where u is the Fourier transform of the NFW halo density profile \citep{Navarro1997}.\\

We fitted the measured correlation functions using a Monte Carlo Markov Chain Metropolis Hastings procedure. We used the same flat priors as in \citet{Coupon2012}. We used only the angular distance larger than 3" to avoid any bias caused by an incorrect deblending. We did not used the number of galaxies ($\bar{n}_{\rm gal}$) as a constraint, since we considered only sub-samples of the full galaxy population at z$\sim$2 (see discussion Sect.\,\ref{sect:validity}). Taking this constraint into account (as e.g. in \citet{Wetzel2013}) would request some extra hypotheses and a large number of extra free parameters, which are too hard to constrain at high redshift. The results could also be potentially biased by the choices of parametric forms used to describe the halo occupation by the various populations. Nevertheless, we a posteriori compared our results with the measurements based on the abundance matching technique and found a good agreement (see Sect.\,\ref{sect:mstarmh}). Strong degeneracies exists between the various HOD parameters. For this reasons, we consider in this paper only the mean host halo mass mass ($\langle M_h \rangle$) defined as
\begin{equation}
\langle M_h \rangle = \int M_h \frac{d^2N}{d\textrm{log}(M_h) dV} \frac{N_c +N_s}{\bar{n}_{\rm gal}} dM_h.
\end{equation}
At each step of the MCMC, we save this quantity, which is derived from the 5 HOD parameters, in order to build the confidence region. However, the probability distribution of $\langle M_h \rangle$ is not flat if we pick random vectors in the 5-dimension HOD parameter space following the priors. For this reason, we divided the probability distribution found by MCMC by the one recovered drawing 100\,000 random sets of HOD parameters.\\

\subsubsection{Using the large-scale clustering}

\label{sect:lsbias}

Sub-samples requiring a PACS selection contain in general a small number of objects and their clustering cannot be estimated accurately at small scale because of the size of the PACS PSF (12" at 160\,$\mu$m). In these particular cases, we used only the large scales ($>$30", corresponding to 0.8\,Mpc comoving at z=2), where the 1-halo term can be neglected, to constraint the effective linear bias of these populations of galaxies. This bias is then converted into halo mass assuming the M$_h$-bias relation of \citet{Tinker2008}. If we assume a constant bias ($b_{\rm pop}$) over the full redshift range, the 2-halo term is then given by
\begin{equation}
w(\theta) = b_{\rm pop}^2 \times w_{\rm DM, \rm pop}
\end{equation}
with 
\begin{equation}
w_{\rm DM, pop} = \frac{\int_z \left ( \frac{dN_{\rm pop}}{dz} \right )^2 \int_{k} \frac{k}{2 \pi} P_{\rm lin}(k,z)J_0(k D_c \theta) \, dz \, dk}{\left ( \int_z \frac{dN_{\rm pop}}{dz}  \, dz \right )^2}.
\end{equation}
We first computed $w_{\rm DM, pop}$, then fitted $b_{\rm pop}$ to the measured ACF, and finally determined the halo mass corresponding to this bias.\\

\subsubsection{Validity and consistency of the two approaches}

\label{sect:validity}

\begin{figure}
\centering
\includegraphics{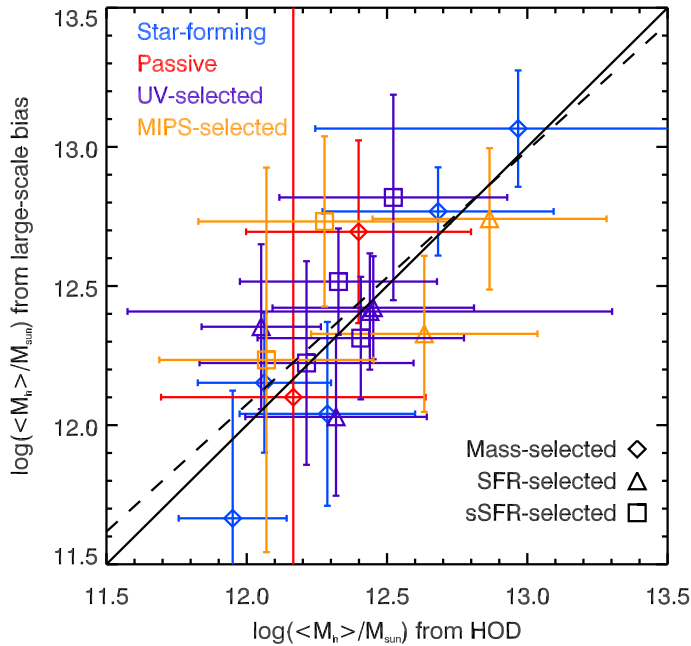}
\caption{\label{fig:consistency} Comparison between the mean host-halo mass derived from the determination of the large-scale ($>$1') bias (see Sect.\,\ref{sect:lsbias}) and from HOD modeling (see Sect.\,\ref{sect:lsbias}). The data points represent all the samples for which the HOD model can be applied, i.e. the non-PACS-selected ones. The solid line is the one-to-one relation and the dashed line the best, linear fit of the data.}
\end{figure}

The two methods presented here have different pros and cons. The HOD approach treats all the scales at the same time and avoids  artificially cutting the data at a scale where the 1-halo term is supposed to be negligible. The method based on the large-scale clustering is simpler, but could be biased by residuals of 1-halo term.  The error bars are similar for both methods, even if the HOD method is based on more data points. This is caused by the degeneracy between the 1-halo and the 2-halo terms, since the large-scale clustering method assumes no 1-halo term. We used HOD modeling when it is possible to constrain sufficiently well the 1-halo, i.e. all our sub-samples except the PACS-selected one.\\

The classical HOD formalism assumes implicitly that the sample is selected by applying a threshold and that the quantity used to define this threshold is correlated with the halo mass. This assumption is not exact for our sub-samples (e.g. SFR-selected sample, sSFR-selected samples, selection by interval and not threshold). We could modify the HOD parametrization, but we cannot know which parametric representation is the most correct before studying the clustering of these populations. However, several problems can happen with the classical HOD applied to our sub-samples. First, our sub-samples occupy only a fraction of the halos and the number of central galaxies never reaches 1. This is not a problem for the HOD modeling if we consider only the clustering, because w($\theta$) stays the same if we multiply $N_c$ and $N_s$ by the same factor $f$. This would be a problem if we had used the $\bar{n}_{\rm gal}$ constraint. In addition, at large halo mass, for star-forming samples, the mean number of central galaxies could tend to zero, because central galaxies of massive halos are often passive. This is compensated artificially by the HOD model, which underestimates the number of satellites to compensate an excess of centrals, since they play a symmetric role in the equations. However, this would be a problem if we had attempted to measure the fraction of satellites, which is studied in another paper using samples selected with a mass threshold (Wolk et al. in prep.).\\
 
We checked the consistency between the two methods comparing the recovered mean halo mass for all the samples used in this paper and for which both methods can be applied. The result is shown Fig.\,\ref{fig:consistency}. We fitted the data with a linear relation and found $\textrm{log}(M_h^\textrm{HOD}/10^{12}\,M_\odot) = (0.91\pm 0.23) \times \textrm{log}(M_h^\textrm{LSB}/10^{12}\,M_\odot) + (0.07\pm 0.12)$ with a reduced $\chi^2$ of 0.59. The slope is thus compatible with unity and the offset is small. If we force the slope at 1, the offset is only 0.02\,dex. There is thus a very good consistency between these two methods, which suffer different systematics. This indicates that at the level of precision reached in this paper, the assumptions we made are reasonable. This is not surprising, because the mean halo mass found by the HOD is strongly related to the large-scale clustering, which is driven by the effective bias of the halos hosting the population (see Eq.\,\ref{eq:hod2h}).\\ 

\section{Relation between halo and stellar mass}

\label{sect:mstarmh}

\begin{figure}
\centering
\includegraphics{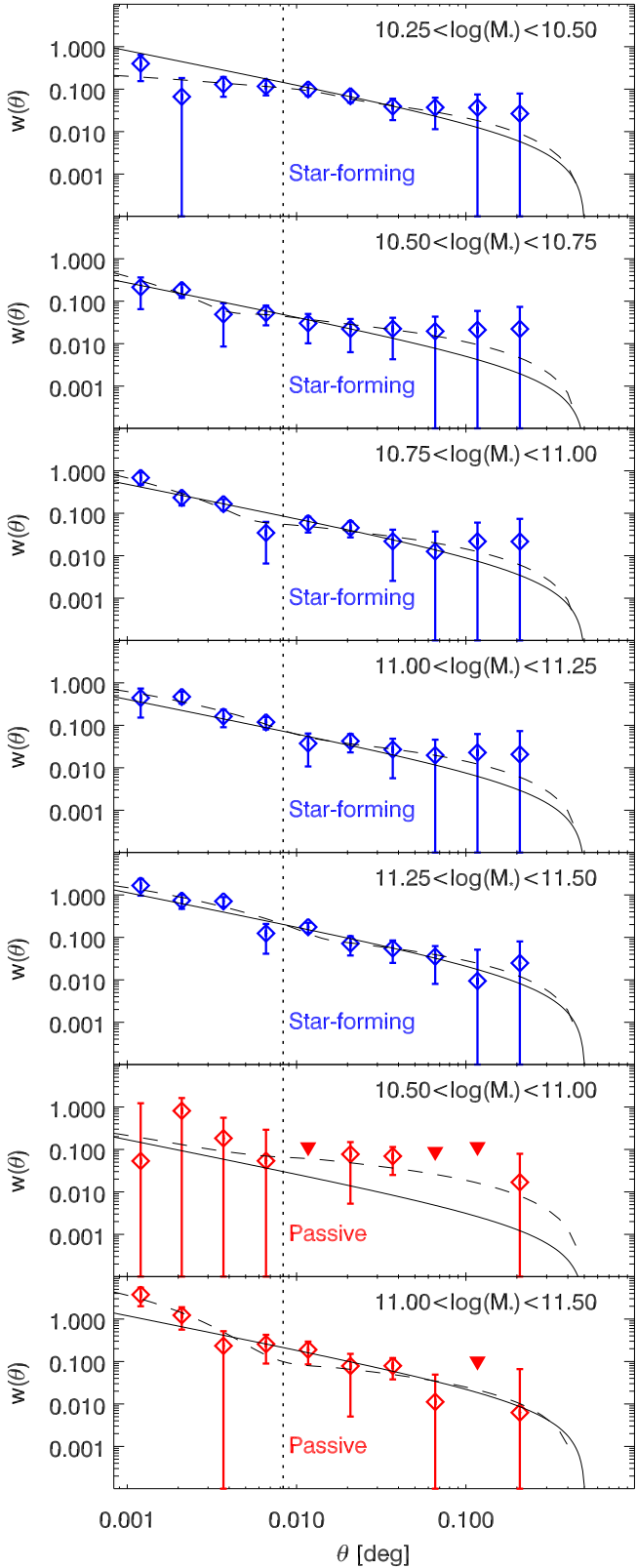}
\caption{\label{fig:clusmass} Auto-correlation function of our various sub-samples of star-forming (blue) and passive (red) galaxies selected by stellar mass. The solid line is the best-fit for the power-law model (fitted only above the 30" limit represented by the vertical dotted line) and the dashed line the best-fit HOD model. The triangles represent the 3\,$\sigma$ upper limits for data points with negative central values, which cannot be represented in a logarithmic plot.}
\end{figure}

\begin{figure}
\centering
\includegraphics{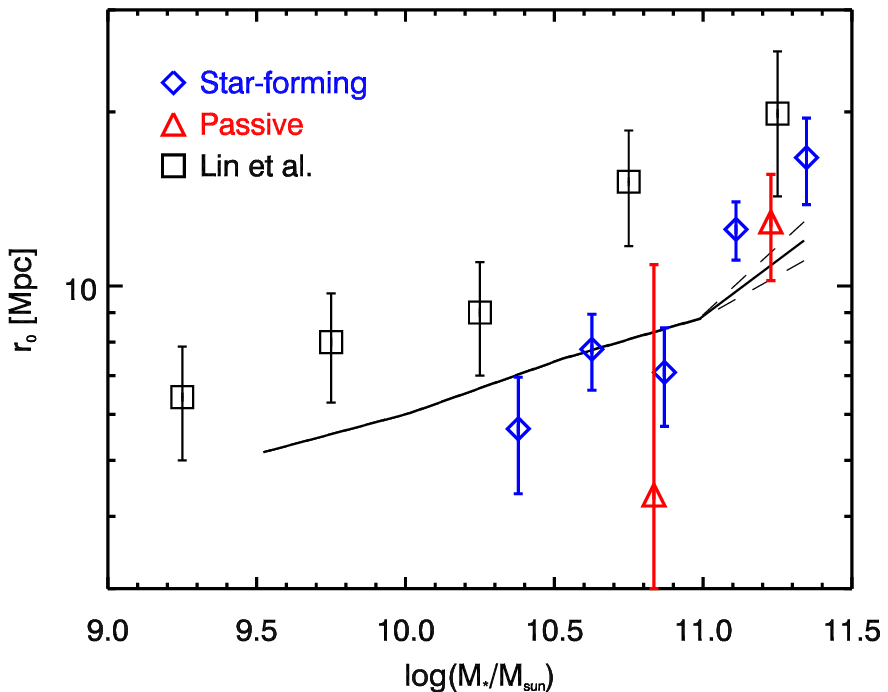}
\includegraphics{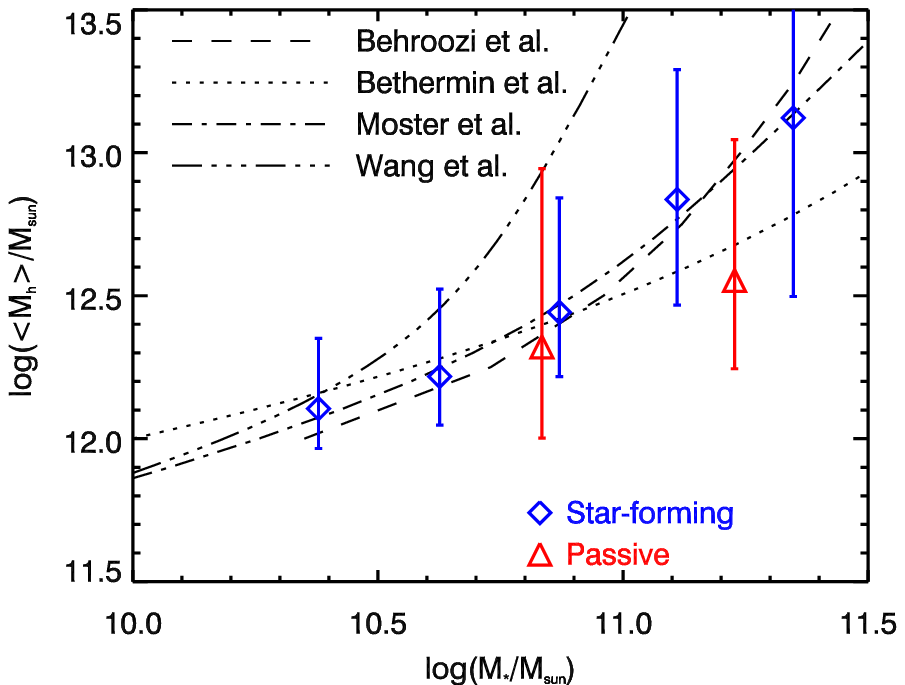}
\caption{\label{fig:resmass} \textit{Upper panel:} Correlation length of star-forming (blue diamonds) and passive (red triangles) galaxies as a function of their stellar mass. For comparison, the data from \citet{Lin2012} are plotted as black squares. The predictions of the \citet{Lagos2011} model is represented with a black solid line, and the 1$\sigma$ confidence region with black dashed line. \textit{Lower panel}: Mean mass of halos, which host star-forming (blue diamonds) and passive (red triangles) galaxies as a function of their mass. For comparison, we over-plotted the estimates performed by abundance matching by \citet{Behroozi2010} (dashed line), \citet{Bethermin2012a} (dotted line), \citet{Moster2010} (dot-dash line), and \citet{Wang2012}.}
\end{figure}

\subsection{Results}

We measured the ACF for various sub-sampled of sBzK and pBzK sorted by stellar mass. For the star-forming galaxies, we used the following bins: $10.25<$log(M$_\star$/M$_\odot$)$<10.50$, $10.50<$log(M$_\star$/M$_\odot$)$<10.75$, $10.75<$log(M$_\star$/M$_\odot$)$<11.00$, $11.00<$log(M$_\star$/M$_\odot$)$<11.25$, and $11.25<$log(M$_\star$/M$_\odot$)$<11.50$. For the passive galaxies, their number density is smaller and we thus used larger mass bins in order to have a sufficient signal: $10.5<$log(M$_\star$/M$_\odot$)$<11.0$, $11.0<$log(M$_\star$/M$_\odot$)$<11.5$. The results are showed in Fig.\,\ref{fig:clusmass}. The results are well fitted by the power-law model at $\theta>30"$ (0.0083$^{\circ}$, vertical dotted line). On smaller scale, we detect some excess due to the 1-halo clustering, especially for passive sub-samples. The HOD model is very flexible and thus nicely fits the data. Fig.\,\ref{fig:resmass} shows the resulting correlation length ($r_0$, upper panel) and mean halo mass (lower panel). These two quantities increase with the stellar mass.\\

We compared our results on the correlation length with the results of \citet{Lin2012} (Fig.\,\ref{fig:resmass}, upper panel). Our results are systematically lower than theirs. However, their analysis was based on the GOODS-N field, which is much smaller than COSMOS (150 versus 7200 arcmin$^2$). They have fitted scales between 3.6" and 6', when we focused on the 30" to 12' range. Consequently, they are more sensitive to the intra-halo clustering, and our analysis is more sensitive to the large-scale clustering. The fact that their values are higher than ours is consistent with the excess of clustering at small scale compared to our power-law fit of the large scales. Their error bars are just slightly larger than ours despite a 13 times smaller field. This is essentially caused by their use of the scales below 30".\\

\subsection{Interpretation}

We found similar clustering lengths to be similar for passive and star-forming galaxies at 1\,$\sigma$. This is in agreement with the results of \citet{Wetzel2013} at z$<$1, who found that the mean halo mass at fixed stellar mass is similar for both populations. This also justifies a posteriori the hypothesis of the same M$_\star$-M$_h$ relation for both populations in the model of \citet{Bethermin2013} linking dark matter halos and infrared galaxy populations. The correlation length is compatible with the prediction of the \citet{Lagos2011} model, contrary to that was claimed by \citet{Lin2012}. This could be caused by the fact that they used very small scale signal for which the 1-halo term could be in excess compared to the power-law approximation and possible problems of deblending. We also compared our M$_\star$-M$_h$ relation with the results of estimates based on abundance matching \citep{Behroozi2010,Moster2010,Bethermin2012a} finding general agreement, except for the case of  \citet{Wang2012}  who predicts larger halo mass for M$_\star > 10^{11}\,$M$_\odot$ galaxies. However, they predict mean stellar mass at fixed halo mass, when we measured the inverse. This nice agreement suggests that the basic hypothesis of the abundance matching, that there is a monotonic relation between halo and stellar mass, is still valid at z=2.\\

\section{Relation between halo mass and star formation rate}

\label{sect:sfrmh}

\begin{figure}
\centering
\includegraphics{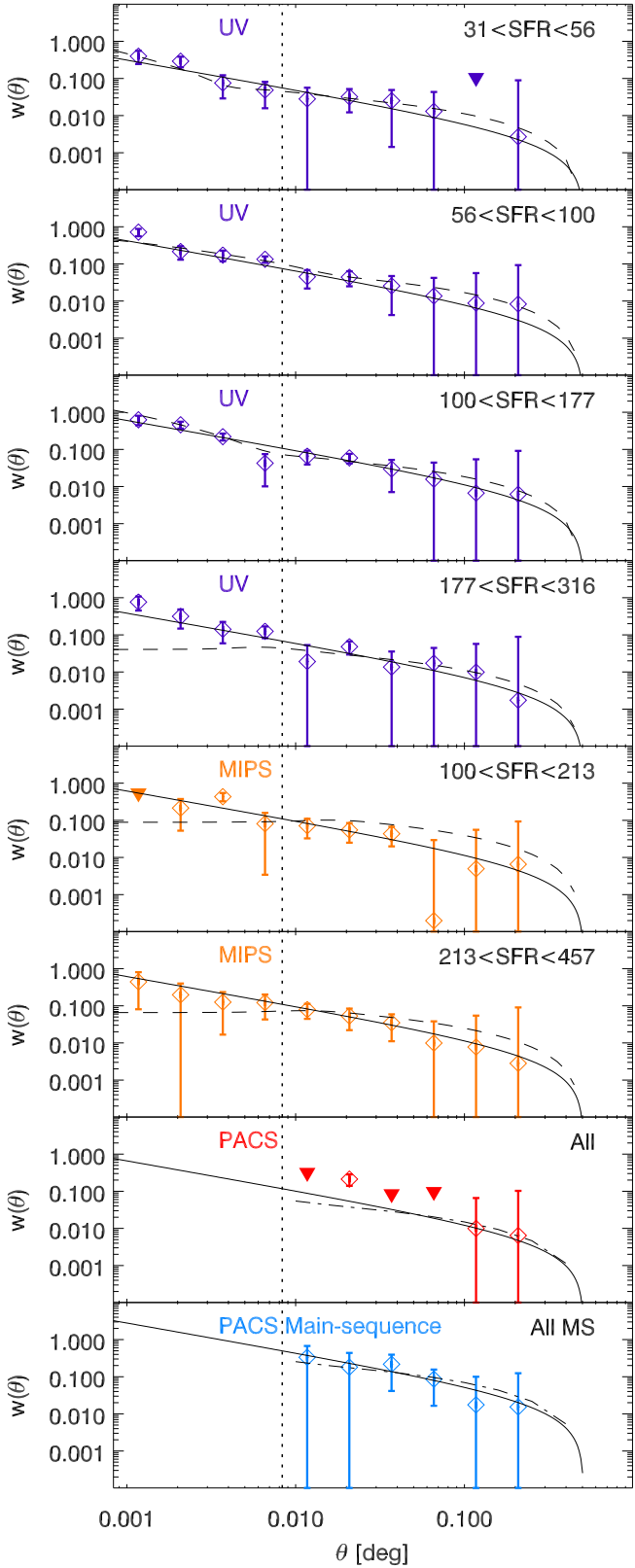}
\caption{\label{fig:clussfr} Auto-correlation function of our various sub-samples as a function of their SFR measured with different tracers (purple for UV, orange for MIPS 24\,$\mu$m, red for PACS, and cyan for PACS main-sequence galaxies). The solid line is the best-fit for the power-law model (fitted only above the 30" limit represented by the vertical dotted line) and the dashed line the best-fit HOD model. The dot-dashed line showed the fit of the linear model for the PACS samples. The triangles represent the 3\,$\sigma$ upper limits for data points with negative central values.}
\end{figure}

\begin{figure}
\centering
\includegraphics{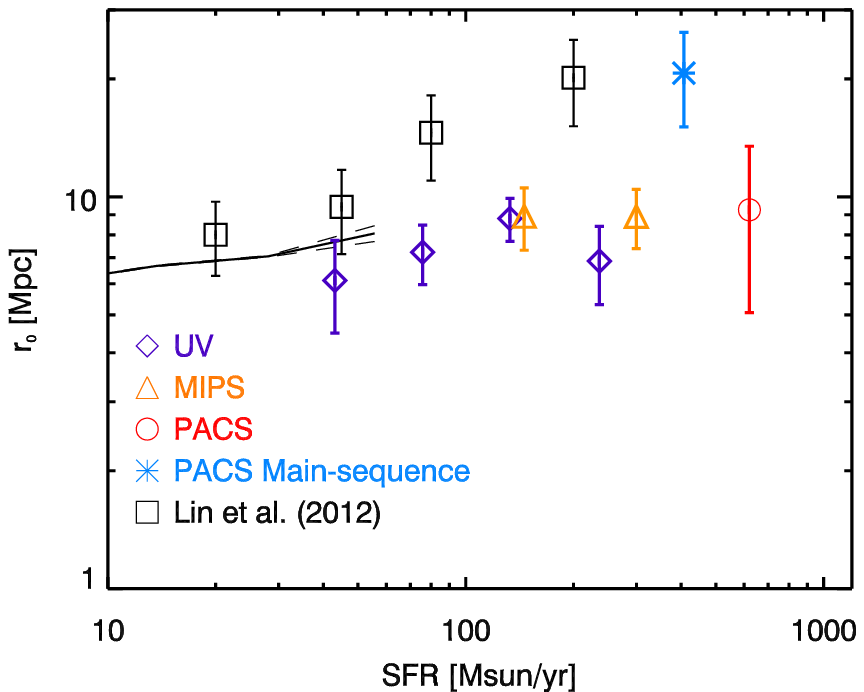}
\includegraphics{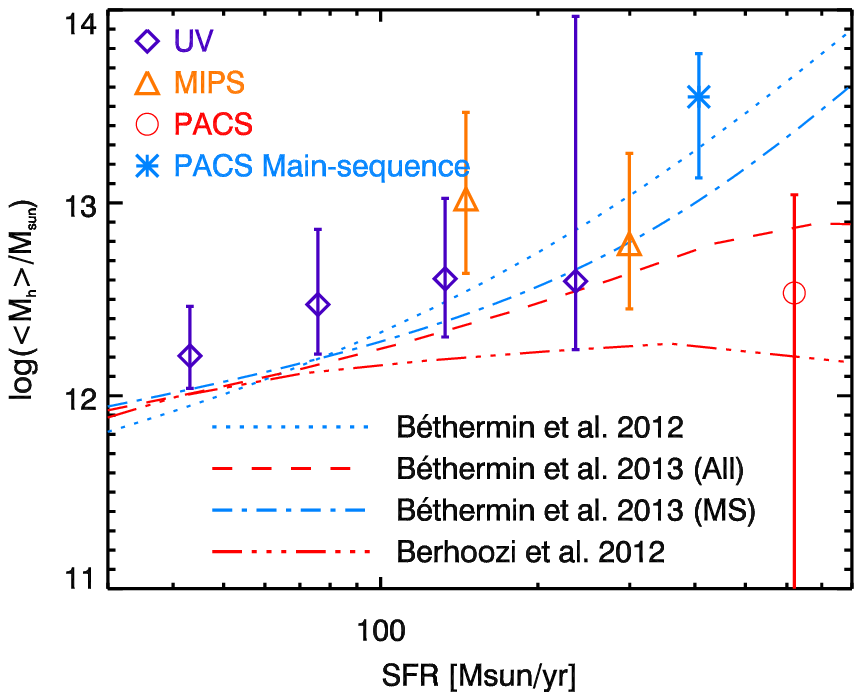}
\caption{\label{fig:ressfr}\textit{Upper panel:} Correlation length of our various sub-samples as a function of their SFR measured at various wavelengths. For comparison, the data from \citet{Lin2012} are plotted with black squares. The predictions of the \citet{Lagos2011} model is represented with a black solid line, and the 1$\sigma$ confidence region with black dashed line. \textit{Lower panel}: Mean mass of halos, which host galaxies as a function of their star formation rate. We over-plotted the estimate performed by abundance matching for main-sequence galaxies of \citet{Bethermin2012a} (dotted line), the predictions of \citet{Bethermin2013} model for main-sequence (dot-dash line) and all (dashed line) galaxies, and the prediction for all galaxies of the \citet{Behroozi2013} model (three-dot-dash line).}
\end{figure}

\subsection{Results}

We then studied the mean halo mass as a function of the SFR. As explained in Sect.\,\ref{sect:sample}, we used by order of priority PACS, MIPS, and UV-derived SFR. We notice that some non-PACS-detected objects have UV-derived and MIPS-derived SFR well above the PACS limit. These sources have an incorrect dust correction for UV SFR and/or an AGN contamination in their 24\,$\mu$m SFR. These sources contaminate the bright flux samples and we excluded them from samples where the SFR limit is sufficiently high that we should be complete using only MIPS and/or PACS detected sources, as they have likely been over-corrected for attenuation. Consequently, we consider several samples: one containing all the sources and using the best available estimator (called hereafter the UV sample), one using only PACS or MIPS by order of priority (called the MIPS sample), and a sample with PACS only. We checked the consistency between the results in the overlapping regions.\\

We used the following bins with similar sizes in logarithmic unit: $\rm 31\,M_\odot.yr^{-1} < SFR<56\,M_\odot.yr^{-1}$, $\rm 56\,M_\odot.yr^{-1} < SFR<100\,M_\odot.yr^{-1}$, $\rm 100\,M_\odot.yr^{-1} < SFR < 177\,M_\odot.yr^{-1}$, $\rm 177\,M_\odot.yr^{-1} < SFR<316\,M_\odot.yr^{-1}$, $\rm 31\,M_\odot.yr^{-1} < SFR<56\,M_\odot.yr^{-1}$ for the UV sample, and $\rm 100\,M_\odot.yr^{-1}<SFR<213\,M_\odot.yr^{-1}$, $\rm 213\,M_\odot.yr^{-1}<SFR<457\,M_\odot.yr^{-1}$ for the MIPS sample. The PACS sample is too small to be split into several sub-samples. We consequently used the full sample. This selection is roughly similar to a SFR$>$200\,M$_\odot$.yr$^{-1}$ selection. Finally, we built a sub-sample of PACS-detected main-sequence galaxies, removing the starbursts from the previous sample. The results are presented in Fig.\,\ref{fig:clussfr}. The data of UV- and MIPS-selected samples  are well fitted by both the power-law and the HOD models. For the PACS-selected samples, we used a power-law fit. The PACS full sample has few objects and is weakly clustered. Consequently the signal is poorly detected and there are a  similar number of positive and negative measurements (3 each). However, the negative points are all close to zero and there is a 2\,$\sigma$ positive outlier. We thus obtain a mean positive signal at $\sim$1\,$\sigma$ by fitting these six points. The clustering signal of PACS-detected, main-sequence galaxies is detected with higher S/N, apparently as  the lower number of objects is largely compensated by a much stronger clustering.\\

Fig.\,\ref{fig:ressfr} presents the correlation length (upper panel) and the mean halo mass (lower panel) as a function of the SFR. Our results disagree with \citet{Lin2012},  (they are looking at smaller scales as explained in the previous section). At SFR$<$200\,M$_\odot$.yr$^{-1}$, we see evidence for a rise of the correlation length and the mean host halo mass with SFR at $\sim$2\,$\sigma$ ($r_0 \propto$ SFR$^{0.32 \pm 0.21}$ and M$_h \propto $ SFR$^{1.2\pm0.6}$). At higher SFR, the data are compatible with a plateau ($r_0 \propto$ SFR$^{0.0 \pm 0.2}$ and M$_h \propto$SFR$^{0.2\pm0.6}$ for SFR$>$100\,M$_\odot$.yr$^{-1}$ data points). This flattening of the SFR-M$_h$ relation is thus significant at only 1.7\,$\sigma$, and future analyses on larger samples will be necessary to confirm wether this trend is real or just a statistical fluctuation. However, if we remove the starbursts from the PACS sample, $r_0$ and $\langle M_h \rangle$ keep rising above 200\,M$_\odot$.yr$^{-1}$. We also compared our results with the model of \citet{Lagos2011}, which agrees with the data around SFR$\sim 50$\,M$_\odot$/yr. Unfortunately, this model predicts very few objects with SFR$>$50\,M$_\odot$/yr, and no prediction can be done above this cut.\\

\subsection{Interpretation}

The evolution of the SFR-$\langle M_h \rangle$ is more difficult to interpret than the M$_\star$-$\langle M_h \rangle$. The increasing mean halo mass with SFR below 200\,M$_\odot$.yr$^{-1}$ is a combined consequence of the monotonic relation between stellar and halo mass, and the correlation between SFR and M$_\star$ for main-sequence galaxies. At higher SFR, the data seem to indicate at 2$\sigma$ the presence of a plateau, or at least a flattening. This regime of SFR is dominated by starburst galaxies, which are above the main-sequence as discussed in \citet{Sargent2012}. In their framework (2SFM), the bright-end of the SFR function is thus caused by galaxies close to the break of the mass function, with a strong excess of sSFR ($>$0.6\,dex). Above 300\,M$_\odot$.yr$^{-1}$, the mean stellar mass no longer increased with SFR. This flattening thus suggests that the stellar mass is better correlated to the halo mass than the SFR. The flattening of the SFR-$\langle M_h \rangle$ relation at the same SFR is also an interesting clue (1.7$\sigma$) of a modification of the star-formation regime at z$\sim$2 around 200\,M$_\odot$.yr$^{-1}$.\\

We compared our results with the predictions of various empirical models (Fig.\,\ref{fig:ressfr}, lower panel). \citet{Bethermin2012a} started from the observed stellar mass function and infrared luminosity function (after removing the starbursts galaxies) and derived the link between SFR and halo mass for main sequence galaxies using an abundance matching technique. Their results agree well with our measurements at SFR$<$200\,M$_\odot$.yr$^{-1}$, where we can assume that the majority of the galaxies lies on the main sequence \citep{Sargent2012}, and with the data point corresponding to PACS-detected main-sequence galaxies. \citet{Bethermin2013} proposed an extended version of this approach taking into account both the starburst and quiescent galaxies. Below 200\,M$_\odot$.yr$^{-1}$, the model provides very similar predictions, if we consider all galaxies or only objects on the main-sequence. At higher SFR, the relation for main-sequence galaxies is steeper and steeper, while the relation for all galaxies flattens. These results agree with the difference of clustering we observe between the all-PACS-detection and PACS-detected, main-sequence samples. Finally, we compared our results with the predictions from \citet{Behroozi2013} model, which was calibrated from the evolution of the stellar mass function. This model predicts a much flatter relation and tends to be slightly lower than the data around 100\,M$_\odot$.yr$^{-1}$ (2$\sigma$ below the MIPS point at 150\,M$_\odot$.yr$^{-1}$).\\

We thus found a typical host halo mass for PACS-detected galaxies of $\sim$10$^{12.5}$\,M$_\odot$. This is about one order of magnitude lower than the measurements of \citet{Magliocchetti2011}, who found 10$^{13.7_{-0.4}^{+0.3}}$\,M$_\odot$ for PACS-detected sources at z$>$1.7. This could be explained by the fact they used the \citet{Mo1996} formalism including scales where the intra-halo clustering is dominant and the potential important cosmic variance caused by the small size of the GOODS-S field. If this is confirmed by future observations on larger fields (e.g. CCAT), this lower value found by our study implies a much lower gap between the typical host halos of local and z$\sim$2 star-forming galaxies mentioned by \citet{Magliocchetti2013}, who consider it as a clue that high-redshift star-forming galaxies do not have the same nature than the local one. Our new results are more in agreement with models based on the idea of a main sequence evolving continuously with redshift, and with a halo mass where star formation is the most efficient evolving very slowly with redshift \citep{Bethermin2012a,Behroozi2013,Bethermin2013}.\\

\section{Relation between halo mass and specific star formation rate}

\label{sect:ssfrmh}

\begin{figure}
\centering
\includegraphics{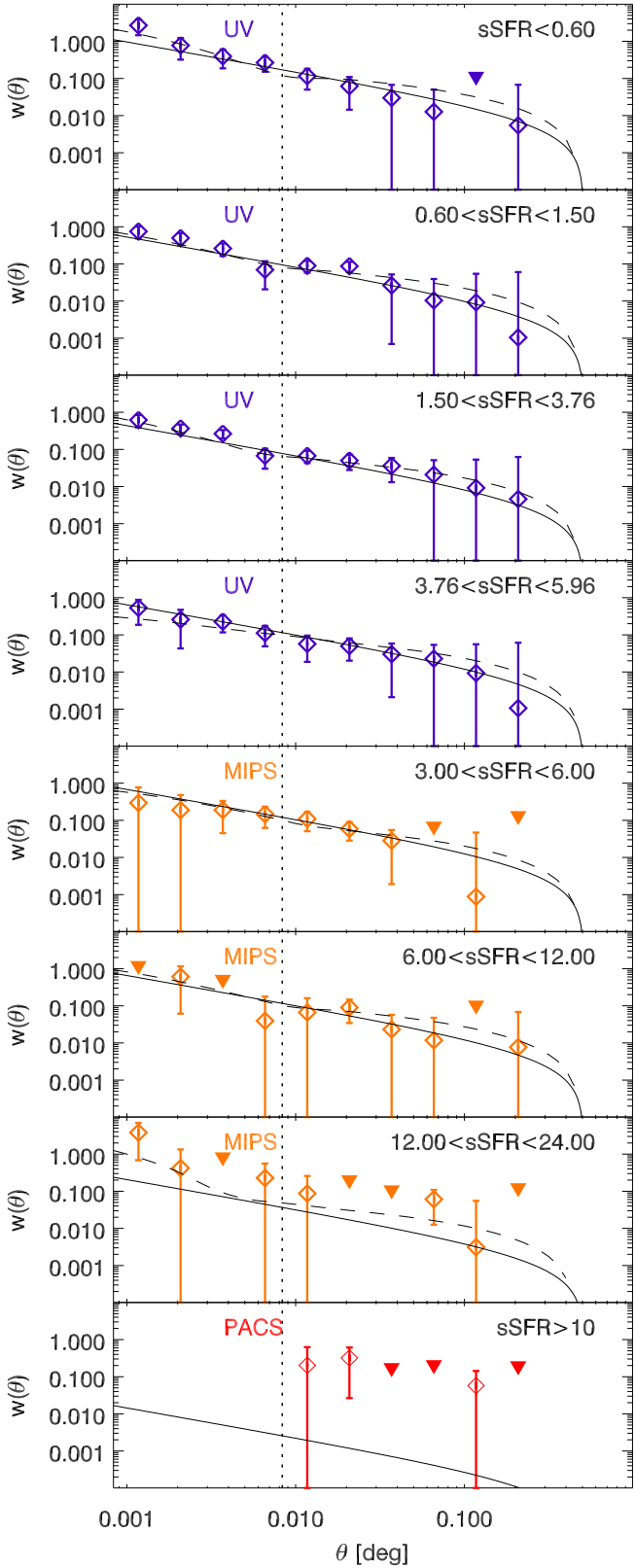}
\caption{\label{fig:clusssfr} Auto-correlation function of our various sub-samples as a function of their sSFR measured at various wavelengths (purple for UV, orange for MIPS 24\,$\mu$m, and red for PACS). The solid line is the best-fit for the power-law model (fitted only above the 30" limit represented by the vertical dotted line) and the dashed line the best-fit HOD model. The triangles represent the 3\,$\sigma$ upper limits for data points with negative central values.}
\end{figure}

\begin{figure}
\centering
\includegraphics{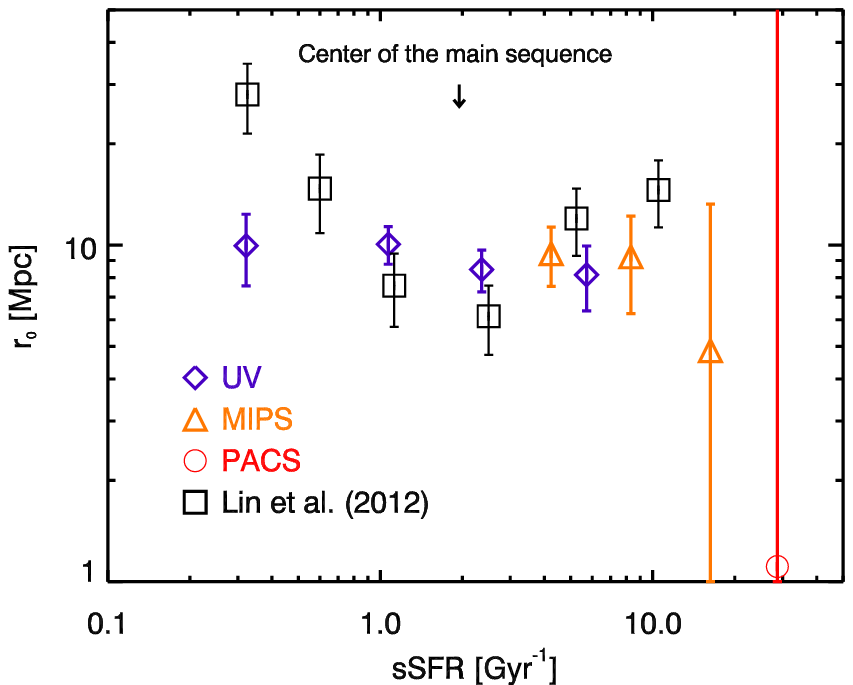}
\includegraphics{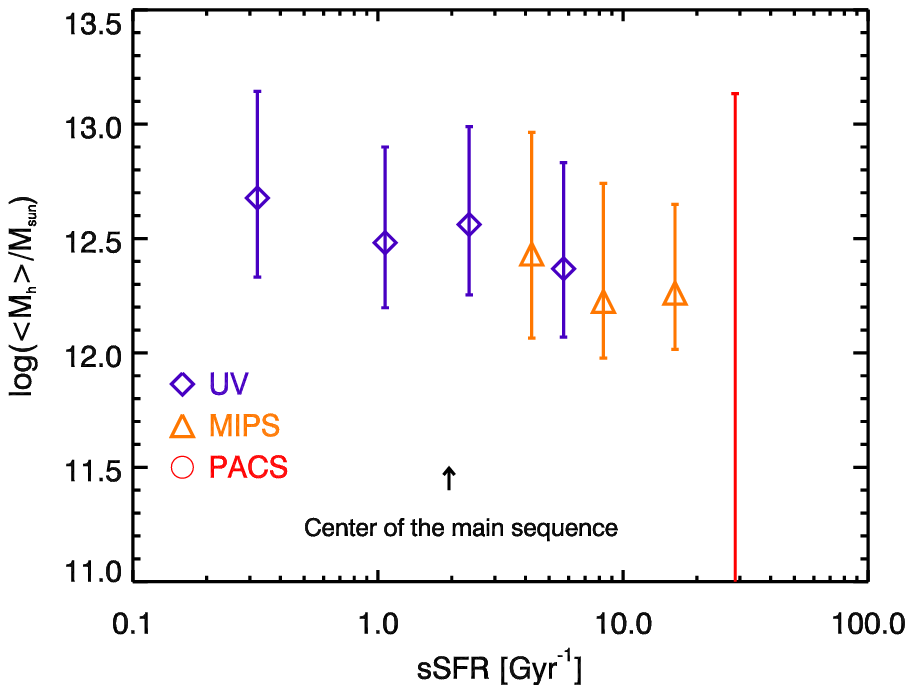}
\caption{\label{fig:resssfr}\textit{Upper panel:} Correlation length of our various sub-samples of log(M$_\star$/M$_\odot$)$>$10.5 galaxies as a function of their sSFR measured at various wavelengths (purple for UV, orange for MIPS 24\,$\mu$m, and red for PACS). For comparison, the data from \citet{Lin2012} are plotted with black squares. \textit{Lower panel}: Mean mass of halos, which host log(M$_\star$/M$_\odot$)$>$10.5 galaxies as a function of their sSFR.}
\end{figure}

\subsection{Results}

Finally, we looked at the clustering as a function of  sSFR for a stellar mass limited sample (log(M$_\star$/M$_\odot$)$>$10.5). The mean log(M$_\star$/M$_\odot$) is quite similar for all the sub-samples and is between 11.75 and 11.88. We used the same UV, MIPS, and PACS samples as in the previous section. The lower sSFR for which the sample is complete is obtained dividing the SFR cut by the mass cut. The sSFR bins are  sSFR$<$0.60\,Gyr$^{-1}$, 0.60\,Gyr$^{-1}$$<$sSFR$<$1.50\,Gyr$^{-1}$, 1.50\,Gyr$^{-1}$$<$sSFR$<$3.76\,Gyr$^{-1}$, 3.76\,Gyr$^{-1}$$<$sSFR$<$5.96\,Gyr$^{-1}$ for the UV samples, 3\,Gyr$^{-1}$$<$sSFR$<$6\,Gyr$^{-1}$, 6\,Gyr$^{-1}$$<$sSFR$<$12\,Gyr$^{-1}$, 12\,Gyr$^{-1}$$<$sSFR$<$24\,Gyr$^{-1}$ for the MIPS sample, and sSFR$>$10\,Gyr$^{-1}$ for the PACS sample. Fig.\,\ref{fig:clusssfr} shows the clustering measurements and their fit by the power-law and the HOD models. Our results exhibit a flat relation between sSFR and the correlation length (Fig.\ref{fig:resssfr}) in disagreement with \citet{Lin2012}, who found minimum of r$_0$ for an sSFR corresponding to the center of the main sequence (see the arrow in the plot corresponding to the position of the center of the main-sequence at z=2 for log(M$_\star$/M$_\odot$)=10.5).\\

\subsection{Interpretation}

\citet{Lin2012} interpreted the excess by the fact that galaxies below and above the main-sequence are associated with dense environments. This disagrees with the flat relation we find between the sSFR and both r$_0$ and M$_h$. Our study is based on more massive galaxies (log(M$_\star$/M$_\odot$)$>$10.5 versus log(M$_\star$/M$_\odot$)$>$9.5), but they checked in their analysis that this trend is not mass dependent. The main cause of the difference is probably the fact that we focused on larger scales ($>$30"), which are essentially associated to the linear clustering and the host halo mass, while they focused on smaller scales (down to 0.3"), which are very sensitive to 1-halo clustering and close environmental effects. To check this hypothesis, we made a new fit using scales down to 2" and found a larger r$_0$ than with the small scale. This new value is at half distance between the \citet{Lin2012} data and our measurements at large scales. At smaller scale and especially below 1", the auto-correlation function in COSMOS is much smaller than their measurements. This could also be partially caused by systematic effects caused by the different deblending methods. In the next section, we will study in detail the clustering of galaxies split into three population (passive, main sequence and starburst) to refine our interpretation of their clustering and understand better this tension.\\

\section{Different clustering properties for main-sequence, starburst and passive populations}

\label{sect:cross}

\subsection{Auto- and cross-correlation functions}

\begin{figure}
\centering
\includegraphics{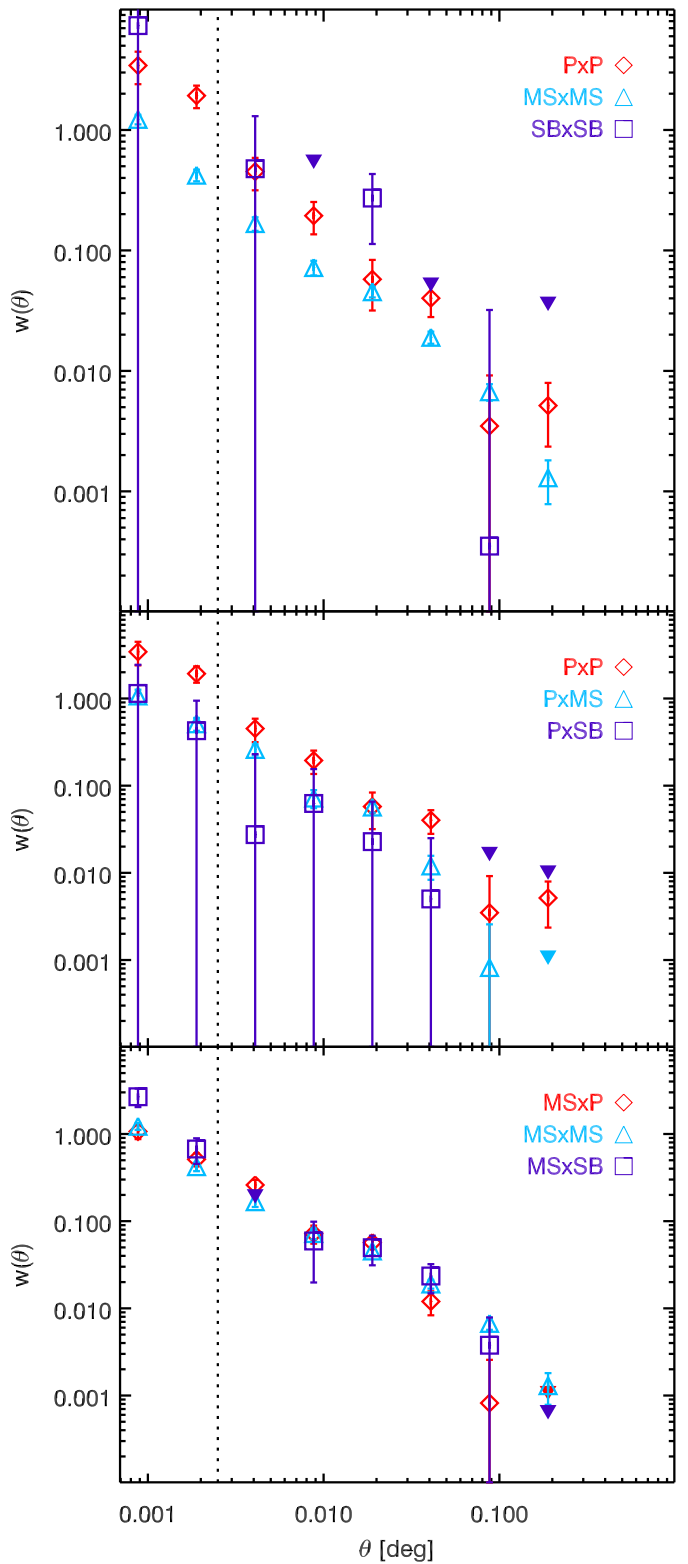}
\caption{\label{fig:crosscorr} \textit{Upper panel:} autocorrelation function of the passive (red diamonds), main-sequence (blue triangles), and starburst (purple squares) log(M$_\star$/M$_\odot$)$>$10.5 galaxies. \textit{Middle panel:} autocorrelation of passive galaxies (red diamonds) compared with the cross-correlation between passive and main-sequence (blue triangles) or starburst (purple squares) galaxies. \textit{Lower panel:} autocorrelation of main-sequence galaxies (blue triangles) compared with the cross-correlation between main-sequence and passive (red triangles) or starburst (purple squares) galaxies. The vertical dotted line indicates the size of the PACS PSF. Below this limit, the reliability of the starburst correlation function is not certain.}
\end{figure}

In this section, we study potential differences between the clustering of galaxies depending of their mode of star formation. We split our z$\sim$2 sample in three mass-selected (M$_\star>10^{10.5}$\,M$_\odot$) sub-samples: the passive (the pBzK sample), main-sequence (sBzk with $\Delta$sSFR$_{\rm MS}$<4), and starburst (PACS-detected sBzK with $\Delta$sSFR$_{\rm MS}$>4) galaxies. The main-sequence and starburst samples have very close mean stellar masses (10.84 versus 10.81). Passive galaxies have a significantly higher mean stellar mass despite a similar mass cut (11.07). This is caused by the steeper slope of the stellar mass function of star-forming galaxies and a slight incompleteness of the sample of passive galaxies at low stellar mass \citep[e.g.][]{Ilbert2013}.\\

The auto-correlation and cross-correlation functions between these populations are shown in Fig.\,\ref{fig:crosscorr}. At large scale ($>$30"), main-sequence and passive galaxies exhibit a similar clustering, but passive galaxies are much more clustered at small scales, as mentioned previously by \citet{McCracken2010}. The origin of this difference is discussed in Sect.\,\ref{sect:smallscale}. The auto-correlation signal from the starburst sample is too weak to draw any conclusion. There is no significant difference between the autocorrelation and the cross-correlation functions between the three subsamples at large scale, except a 2.5$\sigma$ excess of very close ($<$10") pairs of one starburst and one main sequence galaxy. This could be caused by deblending problems. However, we inspected visually the 24\,$\mu$m images and in a majority of the case there is a clear separation between the sources in the close pair, suggesting that this excess is not an artifact. These starbursts could be induced because of the interaction with the more massive neighboring main-sequence galaxies, or the interaction might have higher probability to take place in presence of other close neighbors. At smaller scale, all the cross-correlations provides similar results as the autocorrelation of main-sequence galaxies. This suggests that the excess of clustering of passive galaxies at small scale is probably caused by a mechanism affecting only passive galaxies.\\

\subsection{Similar large-scale clustering properties for all three populations}

To check if the various type of galaxies (passive, main sequence, starburst) are hosted by halos with similar halo masses, when they have similar stellar masses, we measured the large-scale bias of these three populations. The bias of passive and main-sequence galaxies can be measured from the ACF only. The bias of the starbursts is much less constrained. For this reason, we also used the cross-correlation functions between these three populations to derive a much better constraint on the bias of starbursts.\\

Following the Eq.\,\ref{eq:hod2h}, we defined the effective bias of a given population by:
\begin{equation}
b_{\rm eff} = \int_{M_h} \frac{d^2N}{d\textrm{log}(M_h) dV} b(M_h) \frac{N_c + N_s}{\bar{n}_{\rm gal}} dM_h.
\end{equation}
The two-halo term of the cross power-spectrum between two populations A and B is assumed to be: \citep{Cooray2002}
\begin{equation}
P_{A,B} = b_{\rm eff, A} \, b_{\rm eff,B} \, P_{\rm lin}.
\end{equation} 
The two-halo term of the cross-correlation function is thus $\chi_{\rm A,B} =  b_{\rm eff,A} \, b_{\rm eff,B} \, \chi_{\rm DM,A,B}$ with (extended version of Eq.\,\ref{eq:whod})
\begin{equation}
 \chi_{\rm DM,A,B}(\theta)= \frac{\int_z \frac{dN_A}{dz} \frac{dN_B}{dz} \int_{k} \frac{k}{2 \pi} P_{\rm gg}(k,z) J_0(k D_c \theta) \, dz \, dk}{\left ( \int_z \frac{dN_A}{dz}  \, dz \right ) \left ( \int_z \frac{dN_B}{dz}  \, dz \right )},
\end{equation}
We thus determined the value of the three effective bias parameters associated to our three sub-samples by fitting simultaneously the three auto-correlation functions and the three cross-correlation functions. We used only scales larger than 1', where the contribution of the 1-halo term is negligible. The confidence regions of these parameters are determined using an MCMC approach.\\

\begin{figure}
\centering
\includegraphics{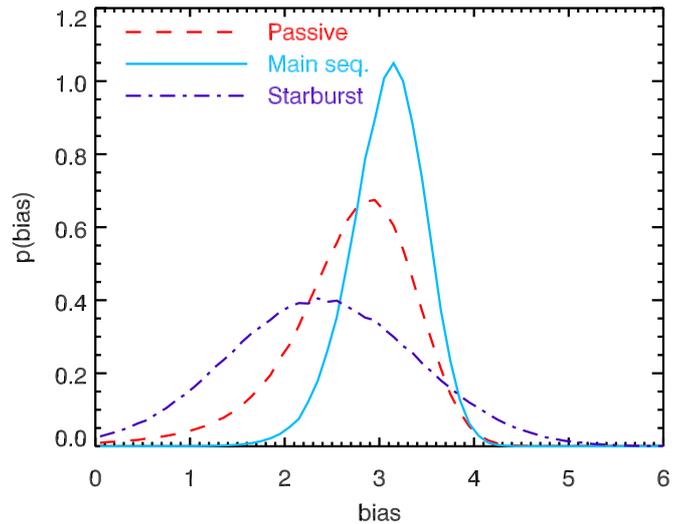}
\caption{\label{fig:bias} Probability distribution of the effective bias of passive (red dashed line), main-sequence (blue solid line), and starburst (purple dot-dashed line) galaxies.}
\end{figure}

Figure\,\ref{fig:bias} shows the probability distribution of the effective bias for each population. The effective bias of the three populations are compatible at 1\,$\sigma$:  2.7$\pm$0.7 for the passive sample, 3.1$\pm$0.4 for the main-sequence sample, and 2.4$\pm$0.9 for the starburst sample. This last measurement would have been impossible without this technique based on the crosscorrelation, since the constraint provided by the ACF on the bias of strabursts is only an upper limit ($<7.0$ at 3$\sigma$). There is a strong disagreement with \citet{Lin2012} on the bias of passive galaxies (7.1$\pm$1.2 in their analysis versus 2.7$\pm$0.7 in our analysis). This means that the bias of passive galaxies at small and large scales is not similar. This result is discussed in Sect.\,\ref{sect:smallscale}.\\

\subsection{The origin of the small-scale clustering of passive galaxies}

\label{sect:smallscale}

\begin{figure}
\centering
\includegraphics{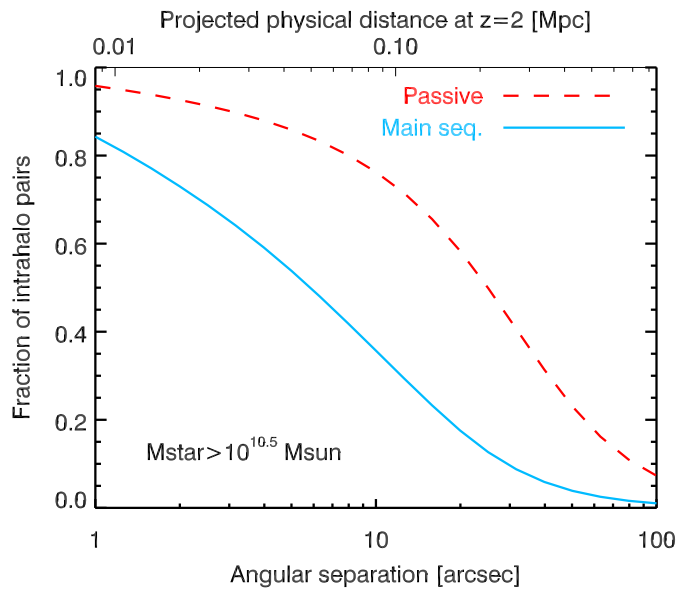}
\caption{\label{fig:physpairs} Fraction of intrahalo pairs as a function of the angular separation for passive (red dashed line) and main-sequence (blue solid line) galaxies.}
\end{figure}

The excess of pairs of passive galaxies at small scale could be explained by the fact that a small fraction of passive galaxies are satellites of central passive galaxies. To test this hypothesis, we selected close pairs of passive galaxies. The fraction of pairs separated by less than $\theta_{\rm max}$, which are associated to galaxies in the same halo (called hereafter intrahalo pairs) can be computed from the autocorrelation function:
\begin{equation}
f_{\rm IH} = \frac{\int_{\theta=0}^{\theta_{\rm max}} \, w_{\rm 1h}(\theta) \theta \, d\theta}{\int_{\theta=0}^{\theta_{\rm max}} \left[1+w(\theta) \right ] \theta \, d\theta}.
\end{equation} 
Figure \,\ref{fig:physpairs} shows this fraction for passive and main-sequence galaxies. This curve was computed from the best HOD fit of the measured ACF of these two populations. We chose to select pairs with a separation smaller than 20". This compromise was chosen to have a balance between the small number of pairs found for very small separations and the low purity for large separations. With this cut, we estimate that 58\,\% of the pairs of passive galaxies are intra-halo pairs.\\

We measured an effective bias squared of all $<20"$ separation pairs of $b_{\rm eff,\,all}^2= 42\pm26$ using only the measurements for $\theta>1'$, where we can make the linear approximation ($w=b_{\rm eff}^2 \, w_{\rm DM}$). We used the barycenter of pairs for our computation of the large scale clustering. This effective bias of all pairs has to be corrected to estimate only the bias associated to real pairs. The total correlation function can be at first order computed from the biases of the intra-halo pairs and pairs caused by chance alinements ($b_{\rm eff,\,IH}$ and $b_{\rm eff,\,CA}$): 
\begin{equation}
w = \left [ f_{\rm IH} b_{\rm eff,\,IH}^2 + 2 \left (1-f_{\rm IH} \right ) b_{\rm eff,\,CA}^2 \right ] w_{\rm DM}.
\end{equation}     
This formula is intuitive if we consider that $w$ is the excess of probability to have a source close to another one compared to the Poisson case. The factor 2 in the second term takes into account the excess of probability to find a source close to both the first and the second sources of pairs caused by chance alinement. This formula neglects the presence of 2-halo pairs with similar redshifts, which are rather negligible for an angular separation lower than 20". We assume that $b_{\rm eff,\,CA}$ is the same as the one of the full population, since only 16\% of our population of passive galaxies is a component of a $<20"$ pair. We found that the bias of intra-halo pairs $b_{\rm eff,\,IH}$ is $7.8_{-3.7}^{+2.4}$. This bias corresponds to a halo mass of 5.5$_{-4.5}^{+5.1}\times10^{13}$\,M$_\sun$.\\

These pairs are thus hosted by structures with a halo mass corresponding to a big group or a small cluster already formed at $z=2$. This type of structures can potentially have sufficiently massive sub-halos to host massive satellite galaxies. We checked that these number agree with the abundance of such halos. For our assumed cosmology we expect a mean of 17 halos more massive than 5.5$\times10^{13}$\,M$_\sun$ between z=1.4 and z=2.5 for our field size, compared to $\sim$80 intracluster pairs based on our estimate of $f_{\rm IH}$. However, an abundance of 80 halos is reached for a cut of 3$\times$10$^{13}$\,M$_\odot$. This value is well inside the 68\% confidence region. There is thus no contradiction between the abundances and the clustering. The origin of the excess of clustering at small scale of passive galaxies at z$\sim$2 is thus probably caused by a small fraction of passive galaxies in massive halos ($\sim 3 \times 10^{13}$\,M$_\odot$). \\

\section{Consistency with X-ray observations}

We have found two types of z$\sim$2 galaxies that appear to be tracing M$_h > 10^{13}$\,M$_\odot$ halos:  massive, strongly-star-forming, PACS-detected, main-sequence galaxies (see Sect.\,\ref{sect:sfrmh}) and close pairs of massive passive galaxies (see Sect.\,\ref{sect:smallscale}). At these halo masses, non negligible X-ray emissions from  hot intra-halo gas might be expected. We thus searched for the X-ray counterparts of our sources to confirm the results of our clustering analysis.\\

\subsection{Direct detections}

The COSMOS field has deep X-ray observations by Chandra and XMM-Newton observatories, allowing us to search for the extended emission down to the level of $8\times 10^{-16}$ ergs s$^{-1}$ cm$^{-2}$ in the 0.5--2 keV band \citep{Finoguenov2007,Leauthaud2010,George2011}. At redshifts above 1, this provides an individual detection of $M_{200}>5 \times 10^{13} M_\odot $ groups (as discussed in other high-z COSMOS group papers, e.g. by \citealt{Onodera2012}).  We find 4 close pairs of massive passive galaxies and 13 PACS-detected  main-sequence sources are coincident with a directly detected extended X-ray emission and could constitute such sources. Most of the sources are however not detected.\\

\subsection{Stacking analysis}

We used the emission-free part of the background-subtracted and exposure corrected X-ray image in the 0.5--2 keV band, with the flux of the detected point source removed to make a stacked flux estimate for the undetected sources. We have produced a further background subtraction refinement, by removing the mean residual flux. The PACS-detected, main-sequence sources produce a $4.4\sigma$ flux enhancement and pairs produce a marginal $2.2\sigma$ flux enhancement at their position, using a 30" aperture. Accounting for the 41\% of the pairs being a random association, this corresponds to an average flux of the group of $2.5 \times 10^{-16}$ ergs s$^{-1}$ cm$^{-2}$ for the pairs and $0.9 \times 10^{-16}$ ergs s$^{-1}$ cm$^{-2}$ for PACS-detected, main-sequence galaxies. Such sources can be individually detected in ultra-deep X-ray surveys, such as CDFS \citep{Finoguenov2014}, and the corresponding total masses of groups are 3.3 and 2.0 $\times 10^{13} M_\odot$, respectively. These are remarkably  similar to the masses inferred from the clustering analysis. The statistical errors on the mean correspond to 0.8 and 0.3 $\times 10^{13} M_\odot$, respectively.\\

\section{Discussion}

\label{sect:discussion}

\subsection{How are galaxies quenched at high redshift?}

\begin{figure}
\centering
\includegraphics{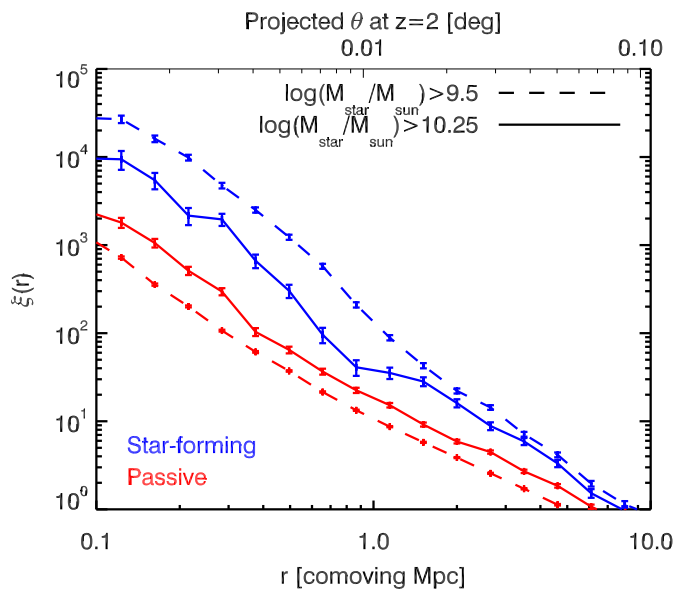}
\caption{\label{fig:gabor}3D autocorrelation function of galaxies at z=2 in the hydrodynamical-simulation of \citet{Gabor2011}. The results for star-forming and passive galaxies are plotted in blue and red, respectively. The results two stellar mass cuts are shown: dashed lines for M$_\star>$10$^{9.5}$\,M$_\odot$ and solid lines for for M$_\star>$10$^{10.25}$\,M$_\odot$.}
\end{figure}

Our measurements indicate that massive (M$_\star > 10^{10.5}$\,M$_\odot$), passive galaxies are in majority central galaxies in $10^{12} - 10^{13}$\,M$_\odot$ halos. However, a small fraction of these passive galaxies are satellites in $>10^{13}$\,M$_\odot$ halos as shown in Sect.\,\ref{sect:smallscale}. For M$_h < 10^{13}$\,M$_\odot$ halos, where sub-halos are not sufficiently massive to host a massive satellite galaxy, the mass quenching is dominant. Our measurements show that this is the dominant process at z$\sim$2 in agreement with \citet{Peng2010}. However, the environmental quenching had apparently already some role at this redshift, as shown by the presence of close pairs of passive galaxies. This role is minor because the number of halos with a sufficiently high mass to host massive galaxies in their sub-structures is much smaller at z$\sim$2 than in the local Universe.\\

We also compared our results with the hydrodynamical simulation of \citet{Gabor2011}. In this simulation the main mechanism of quenching is the formation of a hot halo around massive galaxies preventing the accretion of gas and thus the star formation. This hot atmosphere is both heated by the winds from supernovae and the AGN. Fig.\,\ref{fig:gabor} shows the 3D-correlation function of samples of star-forming and passive galaxies for two various mass cuts ( M$_\star>$10$^{9.5}$\,M$_\odot$ and M$_\star>$10$^{10.25}$\,M$_\odot$). We used this mass cut of 10$^{10.25}$\,M$_\odot$ instead of 10$^{10.5}$\,M$_\odot$ because this simulation produces too few massive galaxies at z$\sim$2 \citep{Gabor2012} and we need a sufficient number of objects to have a reasonable signal. In fact, in this simulation, the clustering is slightly stronger than measured, because the massive galaxies require more massive host halos to form due to a lack of star formation efficiency in this simulation. There is no difference of clustering in the simulation between M$_\star>$10$^{10.25}$\,M$_\odot$ star-forming and passive galaxies at r$>$0.8\,Mpc ($\sim$30" at z=2). This is consistent with our clustering measurements. The model also predicts a large excess of clustering for passive galaxies below 0.8\,Mpc, which is also observed in the data. This hydrodynamical simulation thus predicts the correct trend, even if the mass assembly is not sufficiently quick. The central galaxies begin to be quenched when they reach a high stellar ($\sim 10^{11}$\,M$_\odot$) and halo mass ($\sim 10^{12.5}$\,M$_\odot$). Since the two quantities are correlated, identifying which one is the main driver of the quenching is difficult. The satellite galaxies can be quenched when the hot halo of the central galaxy is formed. This implies that satellite galaxies tends to be more often quenched than centrals at the same mass, and thus the excess of clustering observed at small scales.\\

Our results are thus compatible with the results from the empirical model of \citet{Peng2010} and the hydrodynamical simulation of \citet{Gabor2011}, which both suggests that two different mechanisms of quenching for central and satellite galaxies are already active at z$\sim$2.\\

\subsection{The future of z$\sim$2 populations}

\begin{figure*}
\centering
\begin{tabular}{cc}
\includegraphics{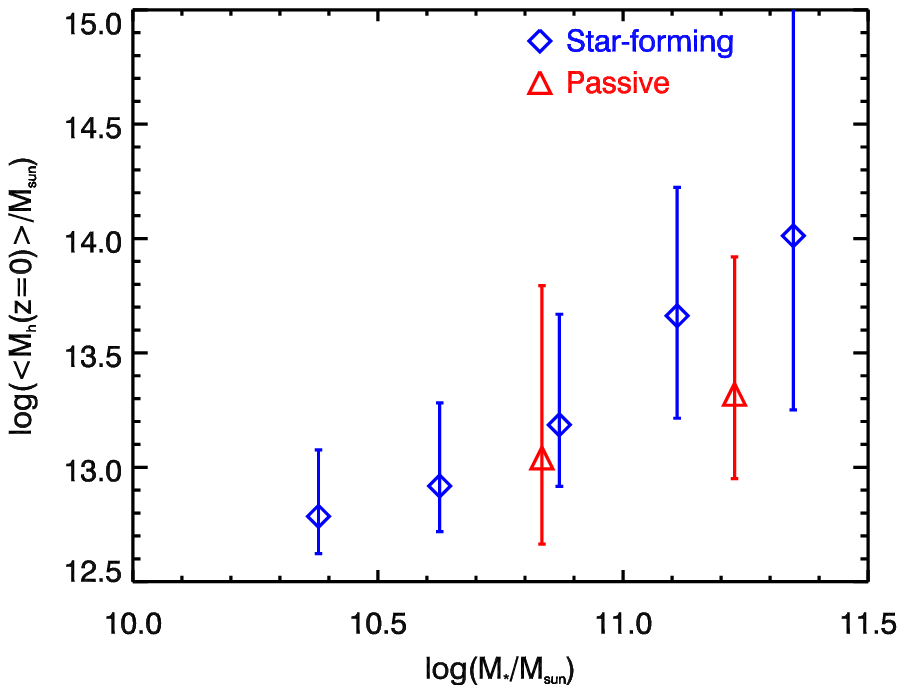} & \includegraphics{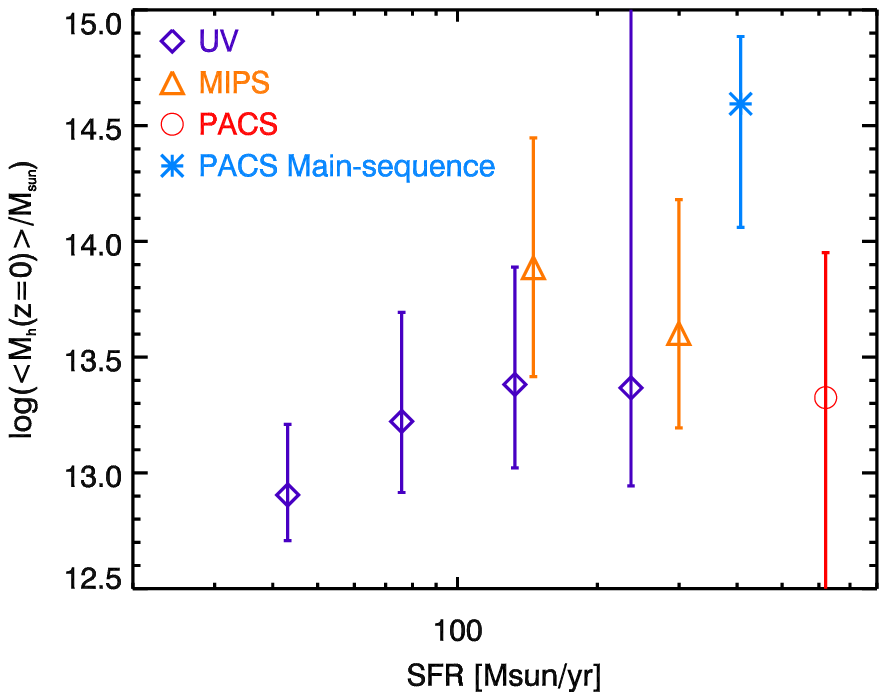}\\
\end{tabular}
\caption{\label{fig:Mhz0} Mean host halo mass at z=0 (extrapolated from z=2 following \citealt{Fakhouri2010}) as a function of the stellar mass (left panel) or SFR (right panel) at z$\sim$2. The symbols are similar as in Fig.\,\ref{fig:resmass} and \ref{fig:ressfr}.}
\end{figure*}

\begin{figure}
\centering
\includegraphics[width=9cm]{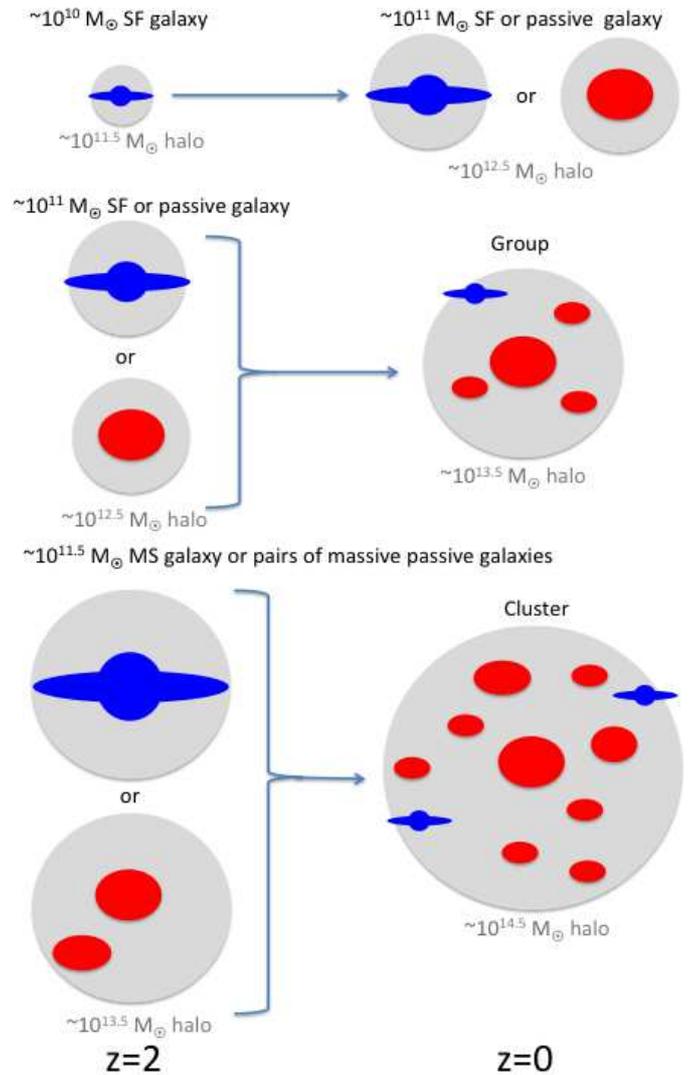}
\caption{\label{fig:cartoon} Cartoon showing a simplified evolution scheme from z=2 to z=0 of galaxy populations studied in this paper.}
\end{figure}

Some interesting insights about galaxy evolution can be obtained studying not only the instantaneous mass of dark matter halos hosting the z=2 galaxies, but also the mass that these structures will have at z=0. We extrapolate the halo mass at z=0 from the halo mass at z=2 using the mean halo growth of \citet{Fakhouri2010}. Figure\,\ref{fig:Mhz0} is similar to Fig.\,\ref{fig:resmass} and \ref{fig:ressfr}, but with the halo mass extrapolated at z=0 instead of the instantaneous halo mass. This allows to connect the z=2 populations studied in our analysis and the descendent population of galaxies at z=0.\\

The main-sequence galaxies with a stellar mass of few 10$^{10}$\,M$_\odot$ and a SFR of few tens of M$_\odot$ per year at z=2 will end up in halos of M$_h \sim 10^{12.5} - 10^{13}$\,M$_\odot$ at z=0. The abundance-matching and weak lensing studies \citep[e.g.][]{Moster2010,Behroozi2010,Leauthaud2012} suggest that these halos host central galaxies with a stellar mass of $\sim$10$^{11}$\,M$_\odot$. At this stellar mass, the mass quenching is efficient according to \citet{Peng2010}, and both passive and star-forming galaxies are observed in the low-z Universe \citep[e.g.][]{Baldry2012,Ilbert2013}. The population of M$_\star \sim$10$^{10}$\,M$_\odot$ sBzK could thus be the progenitor of the most massive field galaxies.\\

The  M$_\star \sim 10^{11}$\,M$_\odot$ sBzKs and pBzKs, hosted by M$_h \sim 10^{12.5}$\,M$_\odot$ z=2 halos, end up in much more massive structures at z=0, with a typical halo mass of $10^{13.5}$\,M$_\odot$ corresponding to big groups and small clusters of galaxies. A significant fraction of these massive galaxies are already quenched at z=2 and the star-forming ones are probably observed just before their quenching, because the mass function of star-forming galaxies at M$_\star >10^{11}$\,M$_\odot$ and especially the massive end is not evolving with redshift as mentioned by \citet{Ilbert2013}.\\

Finally, we identify pairs of passive galaxies (Sect.\,\ref{sect:smallscale}) associated to massive structures formed early. An extrapolation of the growth of their host halos at z=0 gives M$_h = 6.9_{-6.0}^{+9.7}\times10^{14}$\,M$_\odot$. These early-formed groups of passive are thus probably progenitors of the massive clusters in the local Universe, and may be the descendant of the protoclusters of strongly star-forming galaxies observed at $z>4$ \citep[e.g.][]{Daddi2009a,Capak2011}. This simplified evolution picture is summarized by Fig.\,\ref{fig:cartoon}.\\

We can also consider the future of SFR-selected population (see Fig.\,\ref{fig:Mhz0} right). The galaxies with $40<$SFR$<$200\,M$_\odot$.yr$^{-1}$ are progenitors of the central galaxies of groups. The galaxies detected by PACS are expected to be hosted by progenitors of groups as the rest of the population of sBzK if they are episodic starbursts. But PACS-detected main-sequence galaxies are expected to end up in clusters. This suggests that progenitors of clusters can be identified at z=2 using extremely massive star-forming galaxies or groups of massive, passive galaxies. This agrees with \citet{Tanaka2013}, who found a diversity of star formation properties of galaxies in an X-ray-selected groups at z$\sim$1.6. It is still not clear why some of these structures host star-forming galaxies and others passive galaxies.\\

\subsection{The nature of the population of starburst galaxies}

\begin{figure}
\centering
\includegraphics{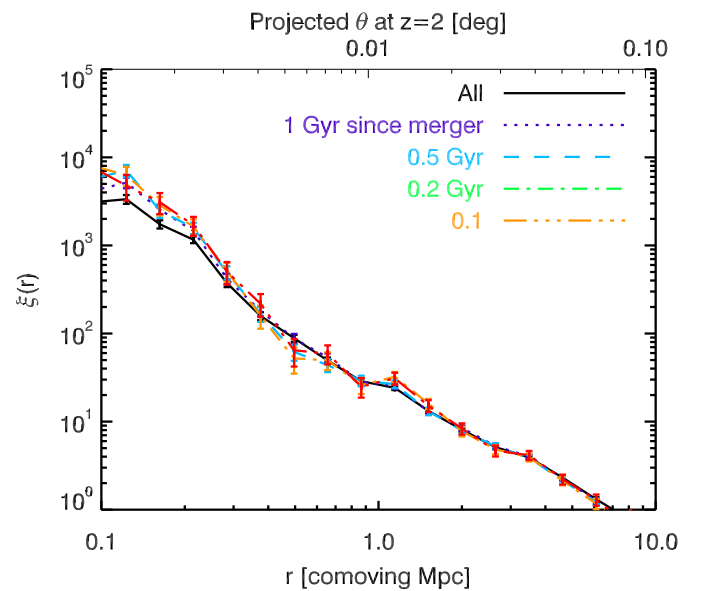}
\caption{\label{fig:gabor_mergers} 3D autocorrelation function of galaxies at z=2 in the \citet{Gabor2011} hydrodynamical simulation as a function of the time since last merger.}
\end{figure}

In the scenario described previously, we neglected the role of the starbursts. As mentioned in the introduction, recent observations favor a scenario where the starbursts host only a minority of the star formation density \citep[$\sim$15\%, e.g.][]{Rodighiero2011,Sargent2012}. However, the mechanisms triggering these violent events are not clearly identified. They could be associated with dense environment such as the group of four SMGs found by \citet{Ivison2013}, or those in \citet{Daddi2009a} and \citet{Chapman2009}. If starbursts are triggered by major mergers, one would naively expect that they are in majority hosted by protoclusters, or some other kind of environmental signature. However, our results show a similar clustering for main-sequence galaxies and starbursts with similar stellar mass, in contradiction with the possibility that starbursts occur in denser environment. In fact, the hydrodynamical simulation of \citet{Gabor2011} shows no excess of clustering above 0.3\,Mpc ($\sim$10" at z=2) for galaxies that merged recently (see Fig.\,\ref{fig:gabor_mergers}). The similar bias thus suggests that M$_\star > 10^{10.5}$ starbursts and main-sequence galaxies are hosted by halos of similar masses. Consequently, starburst episodes do not seem to have a major impact on the M$_\star$-M$_h$ relation in agreement with the idea that they have a minor contribution to the star-formation budget. This finding seems to disagree with the claim of \citet{Michalowski2012} that the stellar mass of SMGs is underestimated and that they lie in fact on the main-sequence, as at fixed SFR they behave as lower-mass objects than main sequence galaxies. On the other hand, we have found some hints for a possible small-scale cross-correlation excess between starbursts and main sequence galaxies that could be pointing to some environmental signature.  The \citet{Gabor2011} model also predicts such a small scale enhancement for recent mergers (Fig.\,\ref{fig:gabor_mergers}).This possibility should be further explored in the future. 

\section{Conclusion}

\label{sect:conclusion}

We measured the clustering of a sample of 25683 star-forming and 2821 passive galaxies in the COSMOS field as a function of their physical properties. This work provided interesting constraints on how the host halo and  environment influence the evolution of galaxies at z$\sim$2, when the cosmic comoving star formation rate density was maximal. Our main findings are as following:
\begin{itemize}
\item We measured the mean host halo mass of z$\sim$2 passive and star-forming galaxies as a function of their stellar mass using only the clustering. Our results agree well with previous estimates based on abundance matching, suggesting that a monotonic relation with scatter between the stellar and the halo mass is already a fair hypothesis at z=2. We also found similar M$_\star$-M$_h$ relations for the two populations.\\

\item We found some clues (2\,$\sigma$) of an increase of the host halo mass with SFR up to 200\,M$_\odot$.yr$^{-1}$, where there is a correlation between SFR, M$_\star$ and M$_h$, and a flattening at higher SFR, where the episodic starbursts with an excess of SFR compared to their stellar mass dominates the population. If we select only main-sequence galaxies, the halo mass continues to rise with SFR above 200\,M$_\odot$.yr$^{-1}$. This transition between the main-sequence and the starburst SFR regime happens at the correct position as predicted in the 2SFM model \citep{Sargent2012,Bethermin2012c}, confirming the relevance of this approach.\\

\item We did not find any difference of large-scale clustering as a function of the sSFR for massive galaxies (M$_\star>$10$^{10.5}$\,M$_\odot$), contrary to \citet{Lin2012} who investigated only smaller scales. This suggests that sSFR does not correlate significantly with the hosting halo mass of the galaxies.\\

\item We confirmed the excess of clustering at small scale ($\theta<30"$) of passive galaxies found by \citet{McCracken2010}. Measuring the large-scale bias of close pairs of passive galaxies, we showed that this is caused by pairs of passive galaxies hosted by the same massive halos ($\sim 3 \times 10^{13}$\,M$_\odot$). This indicates that the environmental quenching is already operating at z$\sim$2, even if mass quenching is dominant. This result is in agreement with the empirical model of \citet{Peng2010} and the hydrodynamical simulation of \citet{Gabor2011}.\\

\item We finally studied the large-scale bias of the population of starburst galaxies using a method based on the angular cross-correlation function between the various populations of galaxies. We found that the bias of starbursts is similar to the one of main-sequence and passive galaxies of the same stellar mass. This suggests that these three populations live in the halos with similar mass, and that starbursts have only a minor role on the assembly of the stellar mass in the halos. Hints of small scale excess are however suggestive of a possible environmental signature.  \\

\item Extrapolating the growth of the halos hosting the populations we studied, we predict that the $\sim 10^{10}$\,M$_\odot$ BzK will end up as massive $\sim 10^{11}$\,M$_\odot$ field galaxies in the local Universe. The $\sim 10^{11}$\,M$_\odot$ massive passive and star-forming BzK lie in progenitors of big groups and small clusters. The future massive clusters can be identified searching for the most massive main-sequence galaxies or groups of massive passive galaxies already formed at z$\sim$2. The halo mass of these z=2 structures was also confirmed by X-ray stacking.\\

\end{itemize}

Thanks to the depth and the area of the COSMOS field from optical to far-infrared, we managed to put first constraints based on clustering measurements on the typical halos hosting the galaxies where the bulk of the star-formation at z$\sim$2 happen. However, some of the results obtained here are only clues (2\,$\sigma$) or weak evidence (3\,$\sigma$). The deep and very large surveys of the next decade in the optical (e.g. LSST), the near-infrared (e.g. \textit{Euclid}), and the sub-millimeter (e.g. CCAT) domains will reduce the uncertainties by typically one order of magnitude because of a similar depth as COSMOS but on fields of $\sim 100$\,deg$^2$ or more. This will allows to study with a much better precision the trends found in this paper and put stronger constraints on the models, but also to have sufficiently large samples to measure the clustering of starbursts at the scale of halos and understand the impact of environmental effect on them.\\

\begin{acknowledgements}
MB, ED, RG, and VS acknowledge the support of the ERC-StG UPGAL 240039 and ANR-08-JCJC-0008 grants. The authors thank Peter Behroozi and Claudia Lagos for providing predictions from their model, and Manuela Magliocchetti and Lihwai Lin for interesting comments.
\end{acknowledgements}

\bibliographystyle{aa}

\bibliography{biblio}

\begin{thebibliography}{92}
\expandafter\ifx\csname natexlab\endcsname\relax\def\natexlab#1{#1}\fi

\bibitem[{{Baldry} {et~al.}(2012){Baldry}, {Driver}, {Loveday}, {Taylor},
  {Kelvin}, {Liske}, {Norberg}, {Robotham}, {Brough}, {Hopkins}, {Bamford},
  {Peacock}, {Bland-Hawthorn}, {Conselice}, {Croom}, {Jones}, {Parkinson},
  {Popescu}, {Prescott}, {Sharp}, \& {Tuffs}}]{Baldry2012}
{Baldry}, I.~K., {Driver}, S.~P., {Loveday}, J., {et~al.} 2012, \mnras, 421,
  621

\bibitem[{{Baugh} {et~al.}(2005){Baugh}, {Lacey}, {Frenk}, {Granato}, {Silva},
  {Bressan}, {Benson}, \& {Cole}}]{Baugh2005}
{Baugh}, C.~M., {Lacey}, C.~G., {Frenk}, C.~S., {et~al.} 2005, \mnras, 356,
  1191

\bibitem[{{Behroozi} {et~al.}(2010){Behroozi}, {Conroy}, \&
  {Wechsler}}]{Behroozi2010}
{Behroozi}, P.~S., {Conroy}, C., \& {Wechsler}, R.~H. 2010, \apj, 717, 379

\bibitem[{{Behroozi} {et~al.}(2013){Behroozi}, {Wechsler}, \&
  {Conroy}}]{Behroozi2013}
{Behroozi}, P.~S., {Wechsler}, R.~H., \& {Conroy}, C. 2013, \apj, 770, 57

\bibitem[{{B{\'e}thermin} {et~al.}(2012{\natexlab{a}}){B{\'e}thermin}, {Daddi},
  {Magdis}, {Sargent}, {Hezaveh}, {Elbaz}, {Le Borgne}, {Mullaney}, {Pannella},
  {Buat}, {Charmandaris}, {Lagache}, \& {Scott}}]{Bethermin2012c}
{B{\'e}thermin}, M., {Daddi}, E., {Magdis}, G., {et~al.} 2012{\natexlab{a}},
  \apjl, 757, L23

\bibitem[{{B{\'e}thermin} {et~al.}(2012{\natexlab{b}}){B{\'e}thermin},
  {Dor{\'e}}, \& {Lagache}}]{Bethermin2012a}
{B{\'e}thermin}, M., {Dor{\'e}}, O., \& {Lagache}, G. 2012{\natexlab{b}}, \aap,
  537, L5

\bibitem[{{B{\'e}thermin} {et~al.}(2012{\natexlab{c}}){B{\'e}thermin}, {Le
  Floc'h}, {Ilbert}, {Conley}, {Lagache}, {Amblard}, {Arumugam}, {Aussel},
  {Berta}, {Bock}, {Boselli}, {Buat}, {Casey}, {Castro-Rodr{\'{\i}}guez},
  {Cava}, {Clements}, {Cooray}, {Dowell}, {Eales}, {Farrah}, {Franceschini},
  {Glenn}, {Griffin}, {Hatziminaoglou}, {Heinis}, {Ibar}, {Ivison},
  {Kartaltepe}, {Levenson}, {Magdis}, {Marchetti}, {Marsden}, {Nguyen},
  {O'Halloran}, {Oliver}, {Omont}, {Page}, {Panuzzo}, {Papageorgiou},
  {Pearson}, {P{\'e}rez-Fournon}, {Pohlen}, {Rigopoulou}, {Roseboom},
  {Rowan-Robinson}, {Salvato}, {Schulz}, {Scott}, {Seymour}, {Shupe}, {Smith},
  {Symeonidis}, {Trichas}, {Tugwell}, {Vaccari}, {Valtchanov}, {Vieira},
  {Viero}, {Wang}, {Xu}, \& {Zemcov}}]{Bethermin2012b}
{B{\'e}thermin}, M., {Le Floc'h}, E., {Ilbert}, O., {et~al.}
  2012{\natexlab{c}}, \aap, 542, A58

\bibitem[{{B{\'e}thermin} {et~al.}(2013){B{\'e}thermin}, {Wang}, {Dor{\'e}},
  {Lagache}, {Sargent}, {Daddi}, {Cousin}, \& {Aussel}}]{Bethermin2013}
{B{\'e}thermin}, M., {Wang}, L., {Dor{\'e}}, O., {et~al.} 2013, ArXiv e-prints

\bibitem[{{Birnboim} {et~al.}(2007){Birnboim}, {Dekel}, \&
  {Neistein}}]{Birnboim2007}
{Birnboim}, Y., {Dekel}, A., \& {Neistein}, E. 2007, \mnras, 380, 339

\bibitem[{{Bouch{\'e}} {et~al.}(2010){Bouch{\'e}}, {Dekel}, {Genzel}, {Genel},
  {Cresci}, {F{\"o}rster Schreiber}, {Shapiro}, {Davies}, \&
  {Tacconi}}]{Bouche2010}
{Bouch{\'e}}, N., {Dekel}, A., {Genzel}, R., {et~al.} 2010, \apj, 718, 1001

\bibitem[{{Burgarella} {et~al.}(2013){Burgarella}, {Buat}, {Gruppioni},
  {Cucciati}, {Heinis}, {Berta}, {B{\'e}thermin}, {Bock}, {Cooray}, {Dunlop},
  {Farrah}, {Franceschini}, {Le Floc'h}, {Lutz}, {Magnelli}, {Nordon},
  {Oliver}, {Page}, {Popesso}, {Pozzi}, {Riguccini}, {Vaccari}, \&
  {Viero}}]{Burgarella2013}
{Burgarella}, D., {Buat}, V., {Gruppioni}, C., {et~al.} 2013, \aap, 554, A70

\bibitem[{{Capak} {et~al.}(2011){Capak}, {Riechers}, {Scoville}, {Carilli},
  {Cox}, {Neri}, {Robertson}, {Salvato}, {Schinnerer}, {Yan}, {Wilson}, {Yun},
  {Civano}, {Elvis}, {Karim}, {Mobasher}, \& {Staguhn}}]{Capak2011}
{Capak}, P.~L., {Riechers}, D., {Scoville}, N.~Z., {et~al.} 2011, \nat, 470,
  233

\bibitem[{{Cattaneo} {et~al.}(2006){Cattaneo}, {Dekel}, {Devriendt},
  {Guiderdoni}, \& {Blaizot}}]{Cattaneo2006}
{Cattaneo}, A., {Dekel}, A., {Devriendt}, J., {Guiderdoni}, B., \& {Blaizot},
  J. 2006, \mnras, 370, 1651

\bibitem[{{Chapman} {et~al.}(2009){Chapman}, {Blain}, {Ibata}, {Ivison},
  {Smail}, \& {Morrison}}]{Chapman2009}
{Chapman}, S.~C., {Blain}, A., {Ibata}, R., {et~al.} 2009, \apj, 691, 560

\bibitem[{{Chapman} {et~al.}(2005){Chapman}, {Blain}, {Smail}, \&
  {Ivison}}]{Chapman2005}
{Chapman}, S.~C., {Blain}, A.~W., {Smail}, I., \& {Ivison}, R.~J. 2005, \apj,
  622, 772

\bibitem[{{Cimatti} {et~al.}(2008){Cimatti}, {Cassata}, {Pozzetti}, {Kurk},
  {Mignoli}, {Renzini}, {Daddi}, {Bolzonella}, {Brusa}, {Rodighiero},
  {Dickinson}, {Franceschini}, {Zamorani}, {Berta}, {Rosati}, \&
  {Halliday}}]{Cimatti2008}
{Cimatti}, A., {Cassata}, P., {Pozzetti}, L., {et~al.} 2008, \aap, 482, 21

\bibitem[{{Conroy} \& {Wechsler}(2009)}]{Conroy2009}
{Conroy}, C. \& {Wechsler}, R.~H. 2009, \apj, 696, 620

\bibitem[{{Cooray} \& {Sheth}(2002)}]{Cooray2002}
{Cooray}, A. \& {Sheth}, R. 2002, \physrep, 372, 1

\bibitem[{{Coupon} {et~al.}(2012){Coupon}, {Kilbinger}, {McCracken}, {Ilbert},
  {Arnouts}, {Mellier}, {Abbas}, {de la Torre}, {Goranova}, {Hudelot}, {Kneib},
  \& {Le F{\`e}vre}}]{Coupon2012}
{Coupon}, J., {Kilbinger}, M., {McCracken}, H.~J., {et~al.} 2012, \aap, 542, A5

\bibitem[{{Cousin} {et~al.}(2013){Cousin}, {Lagache}, {Blaizot}, {Bethermin},
  \& {Guiderdoni}}]{Cousin2013}
{Cousin}, M., {Lagache}, G., {Blaizot}, J., {Bethermin}, M., \& {Guiderdoni},
  B. 2013, sub. to \aap

\bibitem[{{Daddi} {et~al.}(2004){Daddi}, {Cimatti}, {Renzini}, {Fontana},
  {Mignoli}, {Pozzetti}, {Tozzi}, \& {Zamorani}}]{Daddi2004}
{Daddi}, E., {Cimatti}, A., {Renzini}, A., {et~al.} 2004, \apj, 617, 746

\bibitem[{{Daddi} {et~al.}(2009){Daddi}, {Dannerbauer}, {Stern}, {Dickinson},
  {Morrison}, {Elbaz}, {Giavalisco}, {Mancini}, {Pope}, \&
  {Spinrad}}]{Daddi2009a}
{Daddi}, E., {Dannerbauer}, H., {Stern}, D., {et~al.} 2009, \apj, 694, 1517

\bibitem[{{Daddi} {et~al.}(2007){Daddi}, {Dickinson}, {Morrison}, {Chary},
  {Cimatti}, {Elbaz}, {Frayer}, {Renzini}, {Pope}, {Alexander}, {Bauer},
  {Giavalisco}, {Huynh}, {Kurk}, \& {Mignoli}}]{Daddi2007}
{Daddi}, E., {Dickinson}, M., {Morrison}, G., {et~al.} 2007, \apj, 670, 156

\bibitem[{{Daddi} {et~al.}(2010){Daddi}, {Elbaz}, {Walter}, {Bournaud},
  {Salmi}, {Carilli}, {Dannerbauer}, {Dickinson}, {Monaco}, \&
  {Riechers}}]{Daddi2010b}
{Daddi}, E., {Elbaz}, D., {Walter}, F., {et~al.} 2010, \apjl, 714, L118

\bibitem[{{Elbaz} {et~al.}(2011){Elbaz}, {Dickinson}, {Hwang},
  {D{\'{\i}}az-Santos}, {Magdis}, {Magnelli}, {Le Borgne}, {Galliano},
  {Pannella}, {Chanial}, {Armus}, {Charmandaris}, {Daddi}, {Aussel}, {Popesso},
  {Kartaltepe}, {Altieri}, {Valtchanov}, {Coia}, {Dannerbauer}, {Dasyra},
  {Leiton}, {Mazzarella}, {Alexander}, {Buat}, {Burgarella}, {Chary}, {Gilli},
  {Ivison}, {Juneau}, {Le Floc'h}, {Lutz}, {Morrison}, {Mullaney}, {Murphy},
  {Pope}, {Scott}, {Brodwin}, {Calzetti}, {Cesarsky}, {Charlot}, {Dole},
  {Eisenhardt}, {Ferguson}, {F{\"o}rster Schreiber}, {Frayer}, {Giavalisco},
  {Huynh}, {Koekemoer}, {Papovich}, {Reddy}, {Surace}, {Teplitz}, {Yun}, \&
  {Wilson}}]{Elbaz2011}
{Elbaz}, D., {Dickinson}, M., {Hwang}, H.~S., {et~al.} 2011, \aap, 533, A119

\bibitem[{{Fakhouri} \& {Ma}(2010)}]{Fakhouri2010}
{Fakhouri}, O. \& {Ma}, C.-P. 2010, \mnras, 401, 2245

\bibitem[{{Finoguenov} \& {et al.}(2014)}]{Finoguenov2014}
{Finoguenov}, A. \& {et al.} 2014, submitted

\bibitem[{{Finoguenov} {et~al.}(2007){Finoguenov}, {Guzzo}, {Hasinger},
  {Scoville}, {Aussel}, {B{\"o}hringer}, {Brusa}, {Capak}, {Cappelluti},
  {Comastri}, {Giodini}, {Griffiths}, {Impey}, {Koekemoer}, {Kneib},
  {Leauthaud}, {Le F{\`e}vre}, {Lilly}, {Mainieri}, {Massey}, {McCracken},
  {Mobasher}, {Murayama}, {Peacock}, {Sakelliou}, {Schinnerer}, {Silverman},
  {Smol{\v c}i{\'c}}, {Taniguchi}, {Tasca}, {Taylor}, {Trump}, \&
  {Zamorani}}]{Finoguenov2007}
{Finoguenov}, A., {Guzzo}, L., {Hasinger}, G., {et~al.} 2007, \apjs, 172, 182

\bibitem[{{Gabor} \& {Dav{\'e}}(2012)}]{Gabor2012}
{Gabor}, J.~M. \& {Dav{\'e}}, R. 2012, \mnras, 427, 1816

\bibitem[{{Gabor} {et~al.}(2011){Gabor}, {Dav{\'e}}, {Oppenheimer}, \&
  {Finlator}}]{Gabor2011}
{Gabor}, J.~M., {Dav{\'e}}, R., {Oppenheimer}, B.~D., \& {Finlator}, K. 2011,
  \mnras, 417, 2676

\bibitem[{{George} {et~al.}(2011){George}, {Leauthaud}, {Bundy}, {Finoguenov},
  {Tinker}, {Lin}, {Mei}, {Kneib}, {Aussel}, {Behroozi}, {Busha}, {Capak},
  {Coccato}, {Covone}, {Faure}, {Fiorenza}, {Ilbert}, {Le Floc'h}, {Koekemoer},
  {Tanaka}, {Wechsler}, \& {Wolk}}]{George2011}
{George}, M.~R., {Leauthaud}, A., {Bundy}, K., {et~al.} 2011, \apj, 742, 125

\bibitem[{{Goldader} {et~al.}(2002){Goldader}, {Meurer}, {Heckman}, {Seibert},
  {Sanders}, {Calzetti}, \& {Steidel}}]{Goldader2002}
{Goldader}, J.~D., {Meurer}, G., {Heckman}, T.~M., {et~al.} 2002, \apj, 568,
  651

\bibitem[{{Gruppioni} {et~al.}(2013){Gruppioni}, {Pozzi}, {Rodighiero},
  {Delvecchio}, {Berta}, {Pozzetti}, {Zamorani}, {Andreani}, {Cimatti},
  {Ilbert}, {Le Floc'h}, {Lutz}, {Magnelli}, {Marchetti}, {Monaco}, {Nordon},
  {Oliver}, {Popesso}, {Riguccini}, {Roseboom}, {Rosario}, {Sargent},
  {Vaccari}, {Altieri}, {Aussel}, {Bongiovanni}, {Cepa}, {Daddi},
  {Dom{\'{\i}}nguez-S{\'a}nchez}, {Elbaz}, {F{\"o}rster Schreiber}, {Genzel},
  {Iribarrem}, {Magliocchetti}, {Maiolino}, {Poglitsch}, {P{\'e}rez
  Garc{\'{\i}}a}, {Sanchez-Portal}, {Sturm}, {Tacconi}, {Valtchanov},
  {Amblard}, {Arumugam}, {Bethermin}, {Bock}, {Boselli}, {Buat}, {Burgarella},
  {Castro-Rodr{\'{\i}}guez}, {Cava}, {Chanial}, {Clements}, {Conley}, {Cooray},
  {Dowell}, {Dwek}, {Eales}, {Franceschini}, {Glenn}, {Griffin},
  {Hatziminaoglou}, {Ibar}, {Isaak}, {Ivison}, {Lagache}, {Levenson}, {Lu},
  {Madden}, {Maffei}, {Mainetti}, {Nguyen}, {O'Halloran}, {Page}, {Panuzzo},
  {Papageorgiou}, {Pearson}, {P{\'e}rez-Fournon}, {Pohlen}, {Rigopoulou},
  {Rowan-Robinson}, {Schulz}, {Scott}, {Seymour}, {Shupe}, {Smith}, {Stevens},
  {Symeonidis}, {Trichas}, {Tugwell}, {Vigroux}, {Wang}, {Wright}, {Xu},
  {Zemcov}, {Bardelli}, {Carollo}, {Contini}, {Le F{\'e}vre}, {Lilly},
  {Mainieri}, {Renzini}, {Scodeggio}, \& {Zucca}}]{Gruppioni2013}
{Gruppioni}, C., {Pozzi}, F., {Rodighiero}, G., {et~al.} 2013, \mnras, 432, 23

\bibitem[{{Guiderdoni} {et~al.}(1998){Guiderdoni}, {Hivon}, {Bouchet}, \&
  {Maffei}}]{Guiderdoni1998}
{Guiderdoni}, B., {Hivon}, E., {Bouchet}, F.~R., \& {Maffei}, B. 1998, \mnras,
  295, 877

\bibitem[{{Hatton} {et~al.}(2003){Hatton}, {Devriendt}, {Ninin}, {Bouchet},
  {Guiderdoni}, \& {Vibert}}]{Hatton2003}
{Hatton}, S., {Devriendt}, J.~E.~G., {Ninin}, S., {et~al.} 2003, \mnras, 343,
  75

\bibitem[{{Henriques} {et~al.}(2013){Henriques}, {White}, {Thomas}, {Angulo},
  {Guo}, {Lemson}, \& {Springel}}]{Henriques2013}
{Henriques}, B.~M.~B., {White}, S.~D.~M., {Thomas}, P.~A., {et~al.} 2013,
  \mnras, 431, 3373

\bibitem[{{Hopkins} \& {Beacom}(2006)}]{Hopkins2006}
{Hopkins}, A.~M. \& {Beacom}, J.~F. 2006, \apj, 651, 142

\bibitem[{{Hung} {et~al.}(2013){Hung}, {Sanders}, {Casey}, {Lee}, {Barnes},
  {Capak}, {Kartaltepe}, {Koss}, {Larson}, {Le Floc'h}, {Lockhart}, {Man},
  {Mann}, {Riguccini}, {Scoville}, \& {Symeonidis}}]{Hung2013}
{Hung}, C.-L., {Sanders}, D.~B., {Casey}, C.~M., {et~al.} 2013, ArXiv e-prints

\bibitem[{{Ilbert} {et~al.}(2009){Ilbert}, {Capak}, {Salvato}, {Aussel},
  {McCracken}, {Sanders}, {Scoville}, {Kartaltepe}, {Arnouts}, {Le Floc'h},
  {Mobasher}, {Taniguchi}, {Lamareille}, {Leauthaud}, {Sasaki}, {Thompson},
  {Zamojski}, {Zamorani}, {Bardelli}, {Bolzonella}, {Bongiorno}, {Brusa},
  {Caputi}, {Carollo}, {Contini}, {Cook}, {Coppa}, {Cucciati}, {de la Torre},
  {de Ravel}, {Franzetti}, {Garilli}, {Hasinger}, {Iovino}, {Kampczyk},
  {Kneib}, {Knobel}, {Kovac}, {Le Borgne}, {Le Brun}, {F{\`e}vre}, {Lilly},
  {Looper}, {Maier}, {Mainieri}, {Mellier}, {Mignoli}, {Murayama}, {Pell{\`o}},
  {Peng}, {P{\'e}rez-Montero}, {Renzini}, {Ricciardelli}, {Schiminovich},
  {Scodeggio}, {Shioya}, {Silverman}, {Surace}, {Tanaka}, {Tasca}, {Tresse},
  {Vergani}, \& {Zucca}}]{Ilbert2009}
{Ilbert}, O., {Capak}, P., {Salvato}, M., {et~al.} 2009, \apj, 690, 1236

\bibitem[{{Ilbert} {et~al.}(2013){Ilbert}, {McCracken}, {Le Fevre}, {Capak},
  {Dunlop}, {Arnouts}, {Aussel}, {Caputi}, {Comparat}, {Guo}, {Hudelot},
  {Kartaltepe}, {Kneib}, {Krogager}, {Le Floc'h}, {Lilly}, {Mellier},
  {Milvang-Jensen}, {Moutard}, {Onodera}, {Renzini}, {Richard}, {Salvato},
  {Sanders}, {Scoville}, {Silverman}, {Taniguchi}, {Tasca}, {Thomas}, {Toft},
  {Tresse}, {Vergani}, {Wolk}, \& {Zirm}}]{Ilbert2013}
{Ilbert}, O., {McCracken}, H.~J., {Le Fevre}, O., {et~al.} 2013, ArXiv e-prints

\bibitem[{{Ilbert} {et~al.}(2010){Ilbert}, {Salvato}, {Le Floc'h}, {Aussel},
  {Capak}, {McCracken}, {Mobasher}, {Kartaltepe}, {Scoville}, {Sanders},
  {Arnouts}, {Bundy}, {Cassata}, {Kneib}, {Koekemoer}, {Le F{\`e}vre}, {Lilly},
  {Surace}, {Taniguchi}, {Tasca}, {Thompson}, {Tresse}, {Zamojski}, {Zamorani},
  \& {Zucca}}]{Ilbert2010}
{Ilbert}, O., {Salvato}, M., {Le Floc'h}, E., {et~al.} 2010, \apj, 709, 644

\bibitem[{{Ivison} {et~al.}(2013){Ivison}, {Swinbank}, {Smail}, {Harris},
  {Bussmann}, {Cooray}, {Cox}, {Fu}, {Kov{\'a}cs}, {Krips}, {Narayanan},
  {Negrello}, {Neri}, {Pe{\~n}arrubia}, {Richard}, {Riechers}, {Rowlands},
  {Staguhn}, {Targett}, {Amber}, {Baker}, {Bourne}, {Bertoldi}, {Bremer},
  {Calanog}, {Clements}, {Dannerbauer}, {Dariush}, {De Zotti}, {Dunne},
  {Eales}, {Farrah}, {Fleuren}, {Franceschini}, {Geach}, {George}, {Helly},
  {Hopwood}, {Ibar}, {Jarvis}, {Kneib}, {Maddox}, {Omont}, {Scott}, {Serjeant},
  {Smith}, {Thompson}, {Valiante}, {Valtchanov}, {Vieira}, \& {van der
  Werf}}]{Ivison2013}
{Ivison}, R.~J., {Swinbank}, A.~M., {Smail}, I., {et~al.} 2013, \apj, 772, 137

\bibitem[{{Kennicutt}(1998)}]{Kennicutt1998}
{Kennicutt}, Jr., R.~C. 1998, \apj, 498, 541

\bibitem[{{Kere{\v s}} {et~al.}(2005){Kere{\v s}}, {Katz}, {Weinberg}, \&
  {Dav{\'e}}}]{Keres2005}
{Kere{\v s}}, D., {Katz}, N., {Weinberg}, D.~H., \& {Dav{\'e}}, R. 2005,
  \mnras, 363, 2

\bibitem[{{Lagos} {et~al.}(2011){Lagos}, {Lacey}, {Baugh}, {Bower}, \&
  {Benson}}]{Lagos2011}
{Lagos}, C.~D.~P., {Lacey}, C.~G., {Baugh}, C.~M., {Bower}, R.~G., \& {Benson},
  A.~J. 2011, \mnras, 416, 1566

\bibitem[{{Landy} \& {Szalay}(1993)}]{Landy1993}
{Landy}, S.~D. \& {Szalay}, A.~S. 1993, \apj, 412, 64

\bibitem[{{Larson} {et~al.}(2010){Larson}, {Dunkley}, {Hinshaw}, {Komatsu},
  {Nolta}, {Bennett}, {Gold}, {Halpern}, {Hill}, {Jarosik}, {Kogut}, {Limon},
  {Meyer}, {Odegard}, {Page}, {Smith}, {Spergel}, {Tucker}, {Weiland},
  {Wollack}, \& {Wright}}]{Larson2010}
{Larson}, D., {Dunkley}, J., {Hinshaw}, G., {et~al.} 2010, ArXiv e-prints

\bibitem[{{Le Borgne} {et~al.}(2009){Le Borgne}, {Elbaz}, {Ocvirk}, \&
  {Pichon}}]{Le_Borgne2009}
{Le Borgne}, D., {Elbaz}, D., {Ocvirk}, P., \& {Pichon}, C. 2009, \aap, 504,
  727

\bibitem[{{Le Floc'h} {et~al.}(2009){Le Floc'h}, {Aussel}, {Ilbert},
  {Riguccini}, {Frayer}, {Salvato}, {Arnouts}, {Surace}, {Feruglio},
  {Rodighiero}, {Capak}, {Kartaltepe}, {Heinis}, {Sheth}, {Yan}, {McCracken},
  {Thompson}, {Sanders}, {Scoville}, \& {Koekemoer}}]{LeFloch2009}
{Le Floc'h}, E., {Aussel}, H., {Ilbert}, O., {et~al.} 2009, \apj, 703, 222

\bibitem[{{Leauthaud} {et~al.}(2010){Leauthaud}, {Finoguenov}, {Kneib},
  {Taylor}, {Massey}, {Rhodes}, {Ilbert}, {Bundy}, {Tinker}, {George}, {Capak},
  {Koekemoer}, {Johnston}, {Zhang}, {Cappelluti}, {Ellis}, {Elvis}, {Giodini},
  {Heymans}, {Le F{\`e}vre}, {Lilly}, {McCracken}, {Mellier},
  {R{\'e}fr{\'e}gier}, {Salvato}, {Scoville}, {Smoot}, {Tanaka}, {Van
  Waerbeke}, \& {Wolk}}]{Leauthaud2010}
{Leauthaud}, A., {Finoguenov}, A., {Kneib}, J.-P., {et~al.} 2010, \apj, 709, 97

\bibitem[{{Leauthaud} {et~al.}(2012){Leauthaud}, {George}, {Behroozi}, {Bundy},
  {Tinker}, {Wechsler}, {Conroy}, {Finoguenov}, \& {Tanaka}}]{Leauthaud2012}
{Leauthaud}, A., {George}, M.~R., {Behroozi}, P.~S., {et~al.} 2012, \apj, 746,
  95

\bibitem[{{Lee} {et~al.}(2009){Lee}, {Giavalisco}, {Conroy}, {Wechsler},
  {Ferguson}, {Somerville}, {Dickinson}, \& {Urry}}]{Lee2009}
{Lee}, K.-S., {Giavalisco}, M., {Conroy}, C., {et~al.} 2009, \apj, 695, 368

\bibitem[{{Limber}(1953)}]{Limber1953}
{Limber}, D.~N. 1953, \apj, 117, 134

\bibitem[{{Lin} {et~al.}(2012){Lin}, {Dickinson}, {Jian}, {Merson}, {Baugh},
  {Scott}, {Foucaud}, {Wang}, {Yan}, {Yan}, {Cheng}, {Guo}, {Helly}, {Kirsten},
  {Koo}, {Lagos}, {Meger}, {Messias}, {Pope}, {Simard}, {Grogin}, \&
  {Wang}}]{Lin2012}
{Lin}, L., {Dickinson}, M., {Jian}, H.-Y., {et~al.} 2012, \apj, 756, 71

\bibitem[{{Lutz} {et~al.}(2011){Lutz}, {Poglitsch}, {Altieri}, {Andreani},
  {Aussel}, {Berta}, {Bongiovanni}, {Brisbin}, {Cava}, {Cepa}, {Cimatti},
  {Daddi}, {Dominguez-Sanchez}, {Elbaz}, {F{\"o}rster Schreiber}, {Genzel},
  {Grazian}, {Gruppioni}, {Harwit}, {Le Floc'h}, {Magdis}, {Magnelli},
  {Maiolino}, {Nordon}, {P{\'e}rez Garc{\'{\i}}a}, {Popesso}, {Pozzi},
  {Riguccini}, {Rodighiero}, {Saintonge}, {Sanchez Portal}, {Santini}, {Shao},
  {Sturm}, {Tacconi}, {Valtchanov}, {Wetzstein}, \& {Wieprecht}}]{Lutz2011}
{Lutz}, D., {Poglitsch}, A., {Altieri}, B., {et~al.} 2011, \aap, 532, A90

\bibitem[{{Magdis} {et~al.}(2012){Magdis}, {Daddi}, {B{\'e}thermin}, {Sargent},
  {Elbaz}, {Pannella}, {Dickinson}, {Dannerbauer}, {da Cunha}, {Walter},
  {Rigopoulou}, {Charmandaris}, {Hwang}, \& {Kartaltepe}}]{Magdis2012}
{Magdis}, G.~E., {Daddi}, E., {B{\'e}thermin}, M., {et~al.} 2012, \apj, 760, 6

\bibitem[{{Magliocchetti} {et~al.}(2013){Magliocchetti}, {Lapi}, {Negrello},
  {De Zotti}, \& {Danese}}]{Magliocchetti2013}
{Magliocchetti}, M., {Lapi}, A., {Negrello}, M., {De Zotti}, G., \& {Danese},
  L. 2013, ArXiv e-prints

\bibitem[{{Magliocchetti} {et~al.}(2011){Magliocchetti}, {Santini},
  {Rodighiero}, {Grazian}, {Aussel}, {Altieri}, {Andreani}, {Berta}, {Cepa},
  {Casta{\~n}eda}, {Cimatti}, {Daddi}, {Elbaz}, {Genzel}, {Gruppioni}, {Lutz},
  {Magnelli}, {Maiolino}, {Popesso}, {Poglitsch}, {Pozzi}, {Sanchez-Portal},
  {F{\"o}rster Schreiber}, {Sturm}, {Tacconi}, \&
  {Valtchanov}}]{Magliocchetti2011}
{Magliocchetti}, M., {Santini}, P., {Rodighiero}, G., {et~al.} 2011, \mnras,
  416, 1105

\bibitem[{{Magnelli} {et~al.}(2013){Magnelli}, {Popesso}, {Berta}, {Pozzi},
  {Elbaz}, {Lutz}, {Dickinson}, {Altieri}, {Andreani}, {Aussel},
  {B{\'e}thermin}, {Bongiovanni}, {Cepa}, {Charmandaris}, {Chary}, {Cimatti},
  {Daddi}, {F{\"o}rster Schreiber}, {Genzel}, {Gruppioni}, {Harwit}, {Hwang},
  {Ivison}, {Magdis}, {Maiolino}, {Murphy}, {Nordon}, {Pannella}, {P{\'e}rez
  Garc{\'{\i}}a}, {Poglitsch}, {Rosario}, {Sanchez-Portal}, {Santini}, {Scott},
  {Sturm}, {Tacconi}, \& {Valtchanov}}]{Magnelli2013}
{Magnelli}, B., {Popesso}, P., {Berta}, S., {et~al.} 2013, \aap, 553, A132

\bibitem[{{Mandelbaum} {et~al.}(2006){Mandelbaum}, {Seljak}, {Kauffmann},
  {Hirata}, \& {Brinkmann}}]{Mandelbaum2006}
{Mandelbaum}, R., {Seljak}, U., {Kauffmann}, G., {Hirata}, C.~M., \&
  {Brinkmann}, J. 2006, \mnras, 368, 715

\bibitem[{{McCracken} {et~al.}(2010){McCracken}, {Capak}, {Salvato}, {Aussel},
  {Thompson}, {Daddi}, {Sanders}, {Kneib}, {Willott}, {Mancini}, {Renzini},
  {Cook}, {Le F{\`e}vre}, {Ilbert}, {Kartaltepe}, {Koekemoer}, {Mellier},
  {Murayama}, {Scoville}, {Shioya}, \& {Tanaguchi}}]{McCracken2010}
{McCracken}, H.~J., {Capak}, P., {Salvato}, M., {et~al.} 2010, \apj, 708, 202

\bibitem[{{McCracken} {et~al.}(2012){McCracken}, {Milvang-Jensen}, {Dunlop},
  {Franx}, {Fynbo}, {Le F{\`e}vre}, {Holt}, {Caputi}, {Goranova}, {Buitrago},
  {Emerson}, {Freudling}, {Hudelot}, {L{\'o}pez-Sanjuan}, {Magnard}, {Mellier},
  {M{\o}ller}, {Nilsson}, {Sutherland}, {Tasca}, \& {Zabl}}]{McCracken2012}
{McCracken}, H.~J., {Milvang-Jensen}, B., {Dunlop}, J., {et~al.} 2012, \aap,
  544, A156

\bibitem[{{Micha{\l}owski} {et~al.}(2012){Micha{\l}owski}, {Dunlop},
  {Cirasuolo}, {Hjorth}, {Hayward}, \& {Watson}}]{Michalowski2012}
{Micha{\l}owski}, M.~J., {Dunlop}, J.~S., {Cirasuolo}, M., {et~al.} 2012, \aap,
  541, A85

\bibitem[{{Mo} \& {White}(1996)}]{Mo1996}
{Mo}, H.~J. \& {White}, S.~D.~M. 1996, \mnras, 282, 347

\bibitem[{{Moster} {et~al.}(2010){Moster}, {Somerville}, {Maulbetsch}, {van den
  Bosch}, {Macci{\`o}}, {Naab}, \& {Oser}}]{Moster2010}
{Moster}, B.~P., {Somerville}, R.~S., {Maulbetsch}, C., {et~al.} 2010, \apj,
  710, 903

\bibitem[{{Muzzin} {et~al.}(2013){Muzzin}, {Marchesini}, {Stefanon}, {Franx},
  {Milvang-Jensen}, {Dunlop}, {Fynbo}, {Brammer}, {Labb{\'e}}, \& {van
  Dokkum}}]{Muzzin2013}
{Muzzin}, A., {Marchesini}, D., {Stefanon}, M., {et~al.} 2013, \apjs, 206, 8

\bibitem[{{Navarro} {et~al.}(1997){Navarro}, {Frenk}, \& {White}}]{Navarro1997}
{Navarro}, J.~F., {Frenk}, C.~S., \& {White}, S.~D.~M. 1997, \apj, 490, 493

\bibitem[{{Onodera} {et~al.}(2012){Onodera}, {Renzini}, {Carollo},
  {Cappellari}, {Mancini}, {Strazzullo}, {Daddi}, {Arimoto}, {Gobat}, {Yamada},
  {McCracken}, {Ilbert}, {Capak}, {Cimatti}, {Giavalisco}, {Koekemoer}, {Kong},
  {Lilly}, {Motohara}, {Ohta}, {Sanders}, {Scoville}, {Tamura}, \&
  {Taniguchi}}]{Onodera2012}
{Onodera}, M., {Renzini}, A., {Carollo}, M., {et~al.} 2012, \apj, 755, 26

\bibitem[{{Peebles}(1980)}]{Peebles1980}
{Peebles}, P.~J.~E. 1980, {The large-scale structure of the universe}

\bibitem[{{Peng} {et~al.}(2010){Peng}, {Lilly}, {Kova{\v c}}, {Bolzonella},
  {Pozzetti}, {Renzini}, {Zamorani}, {Ilbert}, {Knobel}, {Iovino}, {Maier},
  {Cucciati}, {Tasca}, {Carollo}, {Silverman}, {Kampczyk}, {de Ravel},
  {Sanders}, {Scoville}, {Contini}, {Mainieri}, {Scodeggio}, {Kneib}, {Le
  F{\`e}vre}, {Bardelli}, {Bongiorno}, {Caputi}, {Coppa}, {de la Torre},
  {Franzetti}, {Garilli}, {Lamareille}, {Le Borgne}, {Le Brun}, {Mignoli},
  {Perez Montero}, {Pello}, {Ricciardelli}, {Tanaka}, {Tresse}, {Vergani},
  {Welikala}, {Zucca}, {Oesch}, {Abbas}, {Barnes}, {Bordoloi}, {Bottini},
  {Cappi}, {Cassata}, {Cimatti}, {Fumana}, {Hasinger}, {Koekemoer},
  {Leauthaud}, {Maccagni}, {Marinoni}, {McCracken}, {Memeo}, {Meneux}, {Nair},
  {Porciani}, {Presotto}, \& {Scaramella}}]{Peng2010}
{Peng}, Y.-j., {Lilly}, S.~J., {Kova{\v c}}, K., {et~al.} 2010, \apj, 721, 193

\bibitem[{{Planck Collaboration} {et~al.}(2013){Planck Collaboration}, {Ade},
  {Aghanim}, {Armitage-Caplan}, {Arnaud}, {Ashdown}, {Atrio-Barandela},
  {Aumont}, {Baccigalupi}, {Banday}, \& et~al.}]{Planck_CIB_2013}
{Planck Collaboration}, {Ade}, P.~A.~R., {Aghanim}, N., {et~al.} 2013, ArXiv
  e-prints

\bibitem[{{Rodighiero} {et~al.}(2010){Rodighiero}, {Cimatti}, {Gruppioni},
  {Popesso}, {Andreani}, {Altieri}, {Aussel}, {Berta}, {Bongiovanni},
  {Brisbin}, {Cava}, {Cepa}, {Daddi}, {Dominguez-Sanchez}, {Elbaz}, {Fontana},
  {F{\"o}rster Schreiber}, {Franceschini}, {Genzel}, {Grazian}, {Lutz},
  {Magdis}, {Magliocchetti}, {Magnelli}, {Maiolino}, {Mancini}, {Nordon},
  {Perez Garcia}, {Poglitsch}, {Santini}, {Sanchez-Portal}, {Pozzi},
  {Riguccini}, {Saintonge}, {Shao}, {Sturm}, {Tacconi}, {Valtchanov},
  {Wetzstein}, \& {Wieprecht}}]{Rodighiero2010}
{Rodighiero}, G., {Cimatti}, A., {Gruppioni}, C., {et~al.} 2010, \aap, 518, L25

\bibitem[{{Rodighiero} {et~al.}(2011){Rodighiero}, {Daddi}, {Baronchelli},
  {Cimatti}, {Renzini}, {Aussel}, {Popesso}, {Lutz}, {Andreani}, {Berta},
  {Cava}, {Elbaz}, {Feltre}, {Fontana}, {F{\"o}rster Schreiber},
  {Franceschini}, {Genzel}, {Grazian}, {Gruppioni}, {Ilbert}, {Le Floch},
  {Magdis}, {Magliocchetti}, {Magnelli}, {Maiolino}, {McCracken}, {Nordon},
  {Poglitsch}, {Santini}, {Pozzi}, {Riguccini}, {Tacconi}, {Wuyts}, \&
  {Zamorani}}]{Rodighiero2011}
{Rodighiero}, G., {Daddi}, E., {Baronchelli}, I., {et~al.} 2011, \apjl, 739,
  L40

\bibitem[{{Rodighiero} {et~al.}(2014){Rodighiero}, {Renzini}, {Daddi}, \& {et
  al.}}]{Rodighiero2014}
{Rodighiero}, G., {Renzini}, A., {Daddi}, E., \& {et al.} 2014, sub. to \mnras

\bibitem[{{Salpeter}(1955)}]{Salpeter1955}
{Salpeter}, E.~E. 1955, \apj, 121, 161

\bibitem[{{Sargent} {et~al.}(2012){Sargent}, {B{\'e}thermin}, {Daddi}, \&
  {Elbaz}}]{Sargent2012}
{Sargent}, M.~T., {B{\'e}thermin}, M., {Daddi}, E., \& {Elbaz}, D. 2012, \apjl,
  747, L31

\bibitem[{{Sargent} {et~al.}(2013){Sargent}, {Daddi}, {B{\'e}thermin},
  {Aussel}, {Magdis}, {Hwang}, {Juneau}, {Elbaz}, \& {da Cunha}}]{Sargent2013}
{Sargent}, M.~T., {Daddi}, E., {B{\'e}thermin}, M., {et~al.} 2013, ArXiv
  e-prints

\bibitem[{{Somerville} {et~al.}(2008){Somerville}, {Hopkins}, {Cox},
  {Robertson}, \& {Hernquist}}]{Somerville2008}
{Somerville}, R.~S., {Hopkins}, P.~F., {Cox}, T.~J., {Robertson}, B.~E., \&
  {Hernquist}, L. 2008, \mnras, 391, 481

\bibitem[{{Somerville} \& {Primack}(1999)}]{Somerville1999}
{Somerville}, R.~S. \& {Primack}, J.~R. 1999, \mnras, 310, 1087

\bibitem[{{Tacconi} {et~al.}(2008){Tacconi}, {Genzel}, {Smail}, {Neri},
  {Chapman}, {Ivison}, {Blain}, {Cox}, {Omont}, {Bertoldi}, {Greve},
  {F{\"o}rster Schreiber}, {Genel}, {Lutz}, {Swinbank}, {Shapley}, {Erb},
  {Cimatti}, {Daddi}, \& {Baker}}]{Tacconi2008}
{Tacconi}, L.~J., {Genzel}, R., {Smail}, I., {et~al.} 2008, \apj, 680, 246

\bibitem[{{Tanaka} {et~al.}(2013){Tanaka}, {Finoguenov}, {Mirkazemi}, {Wilman},
  {Mulchaey}, {Ueda}, {Xue}, {Brandt}, \& {Cappelluti}}]{Tanaka2013}
{Tanaka}, M., {Finoguenov}, A., {Mirkazemi}, M., {et~al.} 2013, \pasj, 65, 17

\bibitem[{{Tinker} {et~al.}(2008){Tinker}, {Kravtsov}, {Klypin}, {Abazajian},
  {Warren}, {Yepes}, {Gottl{\"o}ber}, \& {Holz}}]{Tinker2008}
{Tinker}, J., {Kravtsov}, A.~V., {Klypin}, A., {et~al.} 2008, \apj, 688, 709

\bibitem[{{Treister} {et~al.}(2006){Treister}, {Urry}, {Van Duyne},
  {Dickinson}, {Chary}, {Alexander}, {Bauer}, {Natarajan}, {Lira}, \&
  {Grogin}}]{Treister2006}
{Treister}, E., {Urry}, C.~M., {Van Duyne}, J., {et~al.} 2006, \apj, 640, 603

\bibitem[{{Vale} \& {Ostriker}(2004)}]{Vale2004}
{Vale}, A. \& {Ostriker}, J.~P. 2004, \mnras, 353, 189

\bibitem[{{van Kampen} {et~al.}(2005){van Kampen}, {Percival}, {Crawford},
  {Dunlop}, {Scott}, {Bevis}, {Oliver}, {Pearce}, {Kay}, {Gazta{\~n}aga},
  {Hughes}, \& {Aretxaga}}]{VanKampen2005}
{van Kampen}, E., {Percival}, W.~J., {Crawford}, M., {et~al.} 2005, \mnras,
  359, 469

\bibitem[{{Viero} {et~al.}(2013){Viero}, {Wang}, {Zemcov}, {Addison},
  {Amblard}, {Arumugam}, {Aussel}, {B{\'e}thermin}, {Bock}, {Boselli}, {Buat},
  {Burgarella}, {Casey}, {Clements}, {Conley}, {Conversi}, {Cooray}, {De
  Zotti}, {Dowell}, {Farrah}, {Franceschini}, {Glenn}, {Griffin},
  {Hatziminaoglou}, {Heinis}, {Ibar}, {Ivison}, {Lagache}, {Levenson},
  {Marchetti}, {Marsden}, {Nguyen}, {O'Halloran}, {Oliver}, {Omont}, {Page},
  {Papageorgiou}, {Pearson}, {P{\'e}rez-Fournon}, {Pohlen}, {Rigopoulou},
  {Roseboom}, {Rowan-Robinson}, {Schulz}, {Scott}, {Seymour}, {Shupe}, {Smith},
  {Symeonidis}, {Vaccari}, {Valtchanov}, {Vieira}, {Wardlow}, \&
  {Xu}}]{Viero2013}
{Viero}, M.~P., {Wang}, L., {Zemcov}, M., {et~al.} 2013, \apj, 772, 77

\bibitem[{{Wang} {et~al.}(2012){Wang}, {Farrah}, {Oliver}, {Amblard}, {Bock},
  {Conley}, {Cooray}, {Halpern}, {Heinis}, {Ibar}, {Ilbert}, {Ivison},
  {Marsden}, {Roseboom}, {Rowan-Robinson}, {Schulz}, {Smith}, {Viero}, \&
  {Zemcov}}]{Wang2012}
{Wang}, L., {Farrah}, D., {Oliver}, S.~J., {et~al.} 2012, ArXiv e-prints

\bibitem[{{Wetzel} {et~al.}(2013){Wetzel}, {Tinker}, {Conroy}, \& {van den
  Bosch}}]{Wetzel2013}
{Wetzel}, A.~R., {Tinker}, J.~L., {Conroy}, C., \& {van den Bosch}, F.~C. 2013,
  \mnras, 432, 336

\bibitem[{{Williams} {et~al.}(2009){Williams}, {Quadri}, {Franx}, {van Dokkum},
  \& {Labb{\'e}}}]{Williams2009}
{Williams}, R.~J., {Quadri}, R.~F., {Franx}, M., {van Dokkum}, P., \&
  {Labb{\'e}}, I. 2009, \apj, 691, 1879

\bibitem[{{Wuyts} {et~al.}(2011){Wuyts}, {F{\"o}rster Schreiber}, {Lutz},
  {Nordon}, {Berta}, {Altieri}, {Andreani}, {Aussel}, {Bongiovanni}, {Cepa},
  {Cimatti}, {Daddi}, {Elbaz}, {Genzel}, {Koekemoer}, {Magnelli}, {Maiolino},
  {McGrath}, {P{\'e}rez Garc{\'{\i}}a}, {Poglitsch}, {Popesso}, {Pozzi},
  {Sanchez-Portal}, {Sturm}, {Tacconi}, \& {Valtchanov}}]{Wuyts2011a}
{Wuyts}, S., {F{\"o}rster Schreiber}, N.~M., {Lutz}, D., {et~al.} 2011, \apj,
  738, 106

\bibitem[{{Wuyts} {et~al.}(2007){Wuyts}, {Labb{\'e}}, {Franx}, {Rudnick}, {van
  Dokkum}, {Fazio}, {F{\"o}rster Schreiber}, {Huang}, {Moorwood}, {Rix},
  {R{\"o}ttgering}, \& {van der Werf}}]{Wuyts2007}
{Wuyts}, S., {Labb{\'e}}, I., {Franx}, M., {et~al.} 2007, \apj, 655, 51

\bibitem[{{Zheng} {et~al.}(2005){Zheng}, {Berlind}, {Weinberg}, {Benson},
  {Baugh}, {Cole}, {Dav{\'e}}, {Frenk}, {Katz}, \& {Lacey}}]{Zheng2005}
{Zheng}, Z., {Berlind}, A.~A., {Weinberg}, D.~H., {et~al.} 2005, \apj, 633, 791

\end{thebibliography}

\end{document}